\documentclass[twocolumn, prx, superscriptaddress, notitlepage]{revtex4-1}
\usepackage{graphicx}
\usepackage{physics}
\usepackage{float} 
\usepackage{subfigure} 
\usepackage{color}
\usepackage{amsmath}
\usepackage{amsfonts}
\usepackage[normalem]{ulem}
\usepackage{xspace}
\usepackage[dvipsnames]{xcolor}
\usepackage{dcolumn}
\usepackage[breaklinks,colorlinks,linkcolor=blue,citecolor=blue,urlcolor=blue]{hyperref}

\newcommand{\be}{\begin{equation}}
\newcommand{\ee}{\end{equation}}
\newcommand{\bea}{\begin{eqnarray}}
\newcommand{\eea}{\end{eqnarray}}

\begin{document}
\title{Multipolar condensates 
and multipolar Josephson effects}

\author{Wenhui Xu}

 \affiliation{Department of Physics and Astronomy, Purdue University, West Lafayette, IN, 47907, USA}

 \author{Chenwei Lv}

 \affiliation{Department of Physics and Astronomy, Purdue University, West Lafayette, IN, 47907, USA}

\author{Qi Zhou}
\email{zhou753@purdue.edu}
\affiliation{Department of Physics and Astronomy, Purdue University, West Lafayette, IN, 47907, USA}
\affiliation{Purdue Quantum Science and Engineering Institute, Purdue University, West Lafayette, IN 47907, USA}
\date{\today}

\begin{abstract}
When single-particle dynamics are suppressed in certain strongly correlated systems, dipoles arise as  elementary carriers of quantum kinetics. These dipoles can further condense, providing physicists with a rich  realm 
to study fracton phases of matter. Whereas recent theoretical discoveries have shown that an unconventional lattice model may host a dipole condensate as the ground state, fundamental questions arise as to whether dipole condensation is 
a generic phenomenon rather than a specific one unique to a particular model and what new quantum macroscopic phenomena a dipole condensate may bring us with. 
Here, we show that dipole condensates prevail in bosonic systems. Because of a self-proximity effect, where single-particle kinetics inevitably induces a finite order parameter of dipoles, dipole condensation readily occurs in conventional normal phases of bosons. Our findings allow experimentalists to manipulate the phase of a dipole condensate  and deliver dipolar Josephson effects, where supercurrents of dipoles arise in the absence of particle flows. 
The self-proximity effects can also be utilized to produce a generic multipolar condensate. The kinetics of the $n$-th order multipoles unavoidably creates a condensate of the $(n+1)$-th order multipoles, forming a hierarchy of multipolar condensates that will offer physicists a whole new class of macroscopic quantum phenomena.  

\end{abstract}
\maketitle

\section{Introduction}


The pursuit of new quantum matter has been the main theme in modern physics. When interactions produce profound correlations in a many-body system, novel quantum phases arise. Exemplary such phases include but are not limited to superfluids in Helium~\cite{ANDERSON1966}, high-Tc superconductors~\cite{Orenstein2000}, and quantum Hall states~\cite{prange2012quantum}. Recent studies have found a new class of quantum matter, which is constituted by fractons~\cite{Chamon2005, Bravyi2011, Haah2011, Yoshida2013, Fu2015, Fu2016, Nandkishore2019, Pretko2020, Ye2020, Ye2023}. Distinct from conventional particles, fractons are immobile as the movement of a single fracton may 
create extra excitations and thus an additional energy penalty. Bound states of fractons, however, can move and are responsible for the kinetics of the system. The fracton phase of matter has intrigued enthusiastic interest from multiple communities ranging from condensed matter physics to high energy physics and quantum information sciences.  On the one hand, the immobility of fractons may be utilized in quantum computation so as to minimize errors in quantum information processing~\cite{Haah2011, Haah2013, Terhal2015, Williamson2020}. On the other hand, the description of the coupling between fracton and gauge fields requires tensor gauge field theories, a paradigm beyond the conventional vector gauge field theories~\cite{Pretko2017, Chen2018, Barkeshli2018}. It is thus expected that fracton phase of matter will offer physicists a rich playground to explore new physics beyond traditional paradigms.

\begin{figure}[htbp]
\centering  
\subfigure[]{
\label{lineon}
\includegraphics[width=.45\columnwidth]{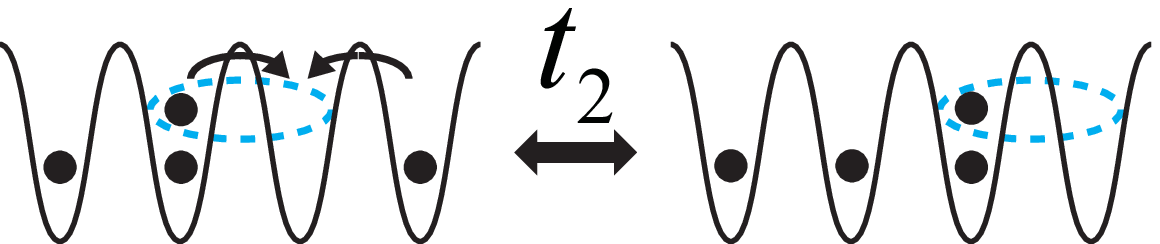}}
\hspace{-5pt}
\subfigure[]{
\label{planon}
\includegraphics[width=.513\columnwidth]{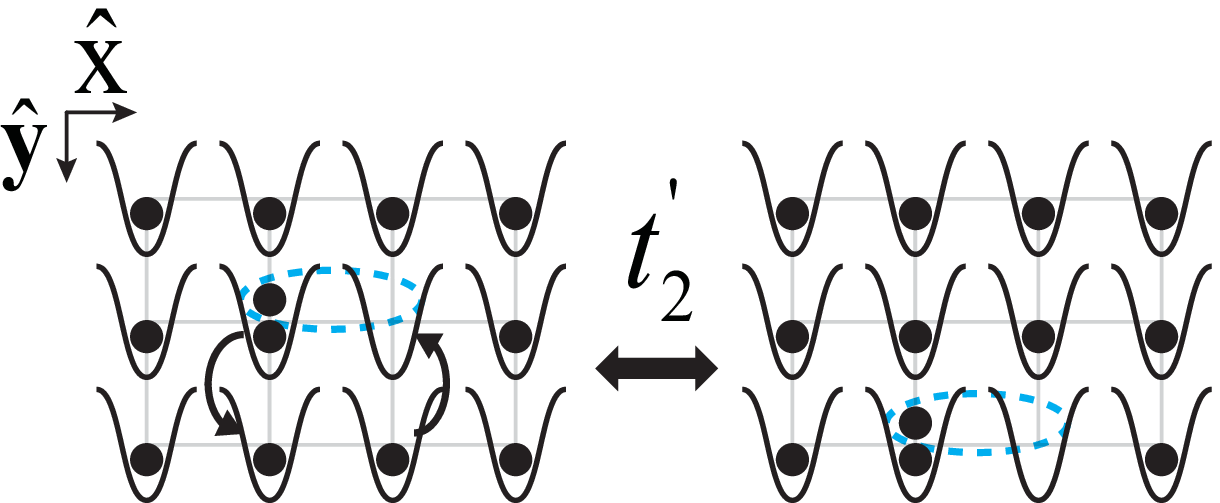}}
\caption{
(a) The correlated tunneling of two particles allows a dipole to move in the direction parallel to its dipole moment. (b)The ring exchange interaction allows a dipole to move in the direction perpendicular to its dipole moment.}
\label{DBHMfig}
\end{figure}

Whereas fracton phase of matter has been explored in a broad range of systems, a recent work has pointed out an elegant scheme to realize such matter using an unconventional bosonic model~\cite{Senthil2022, Senthil2023}. Consider bosons in a tilted lattice, the energy mismatch between the nearest neighbor sites suppresses the tunnelings of single particles and thus delivers fracton physics in a natural manner. A dipole formed by a particle-hole pair, however, can move in the lattice, due to interaction-induced correlated tunnelings. In one dimension, the system is described by the so-called dipolar Bose-Hubbard model(DBHM),
\begin{equation}
    H_{2}=-\sum_i(t_2b_i^2 b_{i-1}^{\dagger}b_{i+1}^{\dagger}+h.c.)+\frac{U}{2}\sum_i n_i(n_i-1)\label{H2}
\end{equation}
where $b_i^\dagger$($b_i$) is the bosonic creation (annihilation) operator at site $i$. $U$ characterizes the usual onsite interaction. The first term in the above equation describes correlated hopping, where two particles initially occupying the same lattice site simultaneously tunnel toward opposite directions. Alternatively, $b^\dagger_i b_{i-1}$ may be regarded as a creation operator for a dipole, and the kinetic energy in Eq.~(\ref{H2}) corresponds to the tunneling of a dipole. As shown in Fig.~\ref{lineon}, a notable feature is that a dipole could only move in the direction parallel to itself and is thus confined in a line. Such a dipole is referred to as a lineon.


The generalization of Eq.~(\ref{H2}) to higher dimensions is straightforward. For instance, in two dimensions, as shown in Fig.~\ref{planon}, a dipole may be able to tunnel in the transverse direction if we consider 
\begin{equation}
    H_{2}'=-\sum_{\bf i}(t_2'b^\dagger_{\bf i+\hat{\bf x}} b^\dagger_{\bf i+\hat{\bf y}} b_{\bf i} b_{\bf i+\hat{\bf x}+\hat{\bf y}}+h.c.)+\frac{U}{2}\sum_{\bf i} n_{\bf i}(n_{\bf i}-1),\label{H2p}
\end{equation}
where ${\bf i}$ is the lattice index, and $\hat{\bf x}$ and $\hat{\bf y}$ are the unit vector in the $x$ and the $y$ directions, respectively. The first term is referred to as the ring exchange interaction. $H_2'$ ensures that a dipole moves in the direction perpendicular to itself. Such a dipole is called a planon. We can certainly include both the ring exchange interaction and the correlated tunneling as that in $H_2$ so as to allow a dipole to move along all directions.  

It has been shown that DBHMs host some intriguing phases as the ground states in different parameter regimes~\cite{Senthil2022, Senthil2023}. 
In one phase,  $\langle b_{\bf i}\rangle\neq 0$, $\langle b^\dagger_{\bf i} b_{{\bf i}+\hat{\bf x}}\rangle\neq 0$, and $\langle b^\dagger_{\bf i} b_{{\bf i}+\hat{\bf y}}\rangle\neq 0$. In other words, both the order parameters of single-particle condensates and dipole condensates are finite. The effective Lagrangian then includes only the quartic term of the phase of the single-particle condensate. Whereas it was previously proposed that such Lagrangian can be accessed using synthetic gauge fields and other methods so as to explore the quantum Lifshitz model~\cite{Zhou2014, Po2015, Zhang2017, Wu2017, Ye2020, Ye2021}, DBHMs allow physicists to study it in a broader parameter regime.

The other phase, where $\langle b_{\bf i}\rangle=0$, $\langle b^\dagger_{\bf i} b_{{\bf i}+\hat{\bf x}}\rangle\neq 0$, and $\langle b^\dagger_{\bf i} b_{{\bf i}+\hat{\bf y}}\rangle\neq 0$, is of particular interest.  
Since the realization of a Bose-Einstein condensate in laboratories~\cite{Anderson1995,Ketterle1995}, 
physicists have been continuously exploring unconventional condensates where single particles do not condense but two or more particles first form a pair or a cluster, and then these pairs or clusters condense~\cite{Pu1998, Svistunov2003, Zhou2011, Kokkelmans2022}. A variety of schemes have been implemented, such as internal degrees of freedom or multiple band effects. However, because of the intrinsic bosonic statistics, single particles naturally form a condensate at sufficiently low temperatures in generic bosonic systems. Experimental efforts in the past many years have proven the difficulty of accessing unconventional condensates in practice.  

The theoretical results of DBHMs suggest a new routine to create dipole condensates in laboratories, for instance, using ultracold atoms in optical lattices. The success of such efforts will provide physicists with many exciting opportunities to explore exotic quantum many-body phenomena. For instance, the condensation of dipoles may gap the tensor gauge fields, as a counterpart of the Higgs mechanism gapping the vector gauge fields~\cite{Chen2018, Barkeshli2018}. 
Based on recent advancements in ultracold atoms experiments~\cite{Dai2017, Yang2020, Bloch2021, Bloch2023}, it is promising that some exotic phenomena of fractons, dipoles, and dipole condensation, as well as their couplings with gauge fields, may be accessed in laboratories in the near future.  

The study of DBHMs imposes some fundamentally important questions. First, whether experimentalists must access the ground state of the DBHMs in Eq.~(\ref{H2}) and Eq.~(\ref{H2p}) so as to create a dipole condensate? If the condensation of dipoles is a more generic phenomenon other than a specific result unique to some particular models, physicists will have a lot more opportunities to explore the fracton phase of matter. Second, in addition to thermal equilibrium, whether a dipole condensate may be created using non-equilibrium quantum dynamics? Studies in the past few decades have shown that non-equilibrium quantum dynamics provide physicists with a much more efficient scheme to access many interesting quantum states than adiabatic processes and other approaches at thermal equilibrium~\cite{Muga2019}. It is desirable to work out some non-equilibrium quantum processes that may deliver dipole condensates as the desired target states. Third, once a dipole condensate is created, what new macroscopic quantum phenomena may be accessible in experiments? Last but not least, in addition to dipole condensates, how to access a generic multipolar condensate and the resultant macroscopic quantum phenomena?



We answer all the above questions in this work.
Our paper is organized as follows. In section II, we discuss the general principle for a dipole condensate to arise.  We also provide some examples of dipole condensates, which turn out to be familiar normal states of bosons. We show that a self-proximity effect makes dipole condensates much more prevalent than one may naively expect. 
 We then present schemes to access dipole condensates using non-equilibrium quantum dynamics in section III. Section IV focuses on a new macroscopic quantum phenomenon, the dipolar Josephson effect where the supercurrent of dipoles emerges 
 after the dipole tunnelings act on dipole condensates with different phases.   In section V, we discuss the hierarchy of multipolar condensates that will allow experimentalists to create a generic multipolar condensate. 
 Section VI includes discussions about experimental realizations of dipole condensates, dipolar Josephson effects, and the generalizations to multipolar condensates. 
 We conclude in section VII with a summary of key findings and an outlook. 


\section{Dipole condensates in BHM}

\subsection{General principles}

\begin{figure}[htbp]
\centering  
\subfigure[]{
\label{phasedia}
\includegraphics[width=.45\columnwidth]{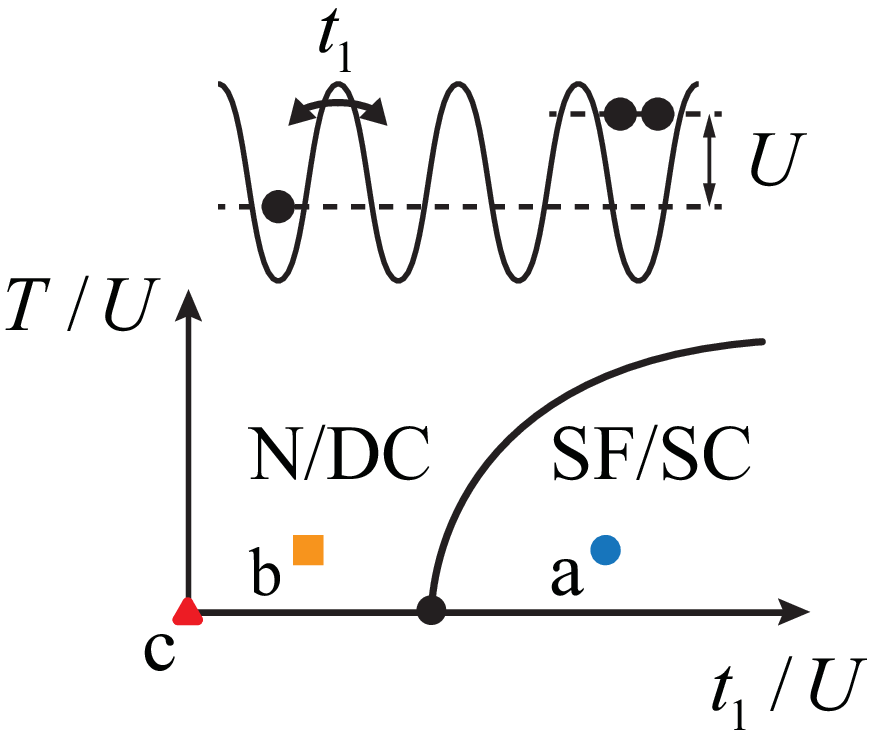}}
\subfigure[]{
\label{correlationlength}
\includegraphics[width=.45\columnwidth]{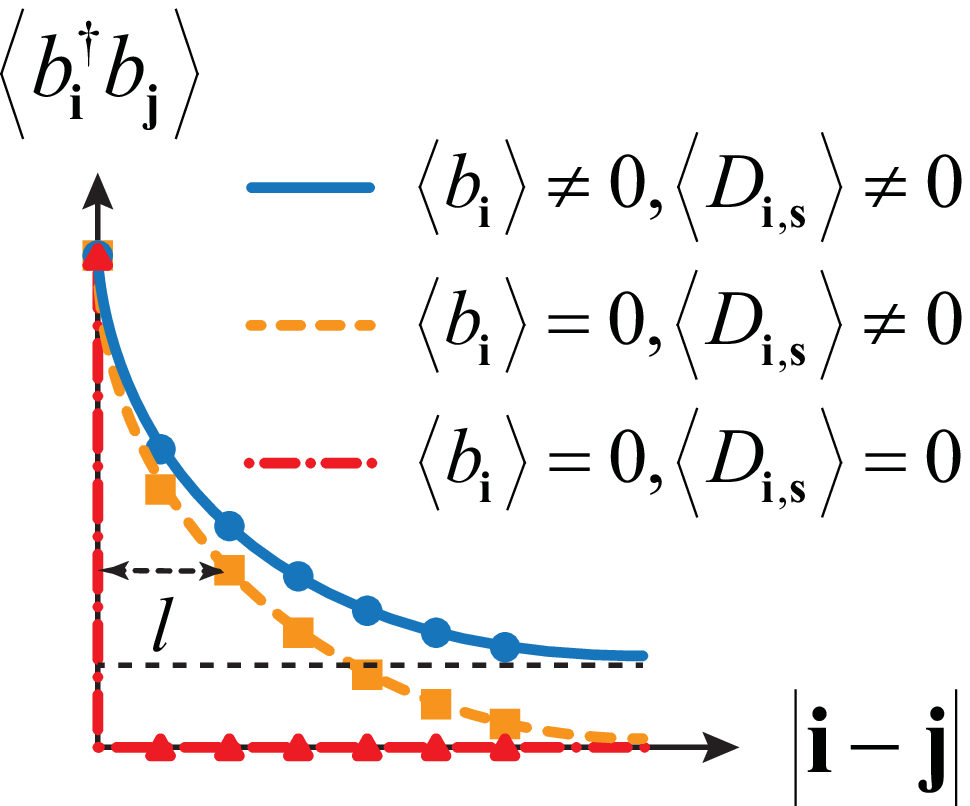}}
\caption{(a) The phase diagram of Bose-Hubbard model. SF/SC denotes the single-particle condensate that may also be referred to as the superfluid phase in 3D. N/DC denotes the normal phase that is also a dipole condensate. (b) Behaviors of the reduced one-body density matrix when the parameters are chosen to be three points in (a). 
The correlation length $l$ vanishes in in the limit of $t_1/U\rightarrow 0$ and $T/U\rightarrow \infty$.}
\label{rho1}
\end{figure}

A conventional condensate is characterized by a finite single-particle order parameter, $\langle b_{\bf i}\rangle\neq 0$. As such, the reduced one-body density matrix $\rho_1({\bf i},{\bf j})\equiv\langle b^\dagger_{\bf i} b_{\bf j}\rangle$ approaches a finite constant when $|{\bf i}-{\bf j}|\rightarrow \infty$, defining the off-diagonal long-range order. When the single-particle condensate vanishes, $\langle b_{\bf i}\rangle=0$, and $\langle b^\dagger_{\bf i} b_{\bf j}\rangle\rightarrow 0$ in the large $|{\bf i}-{\bf j}|$ limit, signifying the vanishing off-diagonal long-range order due to either the interaction effect or thermal fluctuations. Under this situation, the state is often considered as a normal phase of bosons. Nevertheless, short-range correlations may still exist, and  $\langle b^\dagger_{\bf i} b_{\bf j}\rangle$ remains finite for a finite $|{\bf i}-{\bf j}|$, characterized by a finite correlation length $l$, as shown in Fig.~\ref{correlationlength}. In some extreme cases,  $\langle b^\dagger_{\bf i} b_{\bf j}\rangle=n_0\delta_{{\bf i}-{\bf j}}$, where $n_0$ is the particle number per site, short-range correlations also vanish. This can be accessed in the strong interaction limit at  zero temperature, where the ground state becomes a Mott insulator, 
\begin{equation}|{\text{MI}}\rangle=\prod_{\bf i}\frac{b_{\bf i}^{\dagger n_0}}{\sqrt{n_0!}}|0\rangle.
\label{mi}\end{equation}
Alternatively,  the correlation can  be suppressed by thermal fluctuations. At the infinite temperature, the correlation length vanishes, and $\langle b^\dagger_{\bf i} b_{\bf j}\rangle=n_0\delta_{{\bf i}-{\bf j}}$. 

A dipole condensate is characterized by a finite two-body order parameter $\langle D_{{\bf i},{\bf s}}\rangle\neq 0$, where $D_{{\bf i},{\bf s}}\equiv b^\dagger_{\bf i} b_{\bf{i}+\bf{s}}$ and ${\bf{s}}\neq 0$. The order parameter, $\langle b^\dagger_{\bf i} b_{\bf{i}+\hat{x}}\rangle$, considered in DBHMs is a special case where the separation between the particle and hole is one lattice spacing. More generically,  $D_{{\bf i},{\bf s}}$ is the creation operator for a dipole, whose center of mass is located at ${\bf i}+{\bf s}/2$, and ${\bf s}$ is regarded as the relative coordinate of the particle-hole pair. In other words, 
the reduced one-body density matrix captures the order parameter of the dipole condensate. As such, a conventional normal phase of bosons with a finite correlation length $l$ readily acquires a finite dipole order parameter $\langle D_{{\bf i},{\bf s}}\rangle$ for a finite ${\bf s}$, and thus is a dipole condensate. The only exception is  the extreme case where the correlation length $l$ vanishes and $\langle D_{{\bf i},{\bf s}}\rangle$ becomes zero for any finite ${\bf s}$. For any finite $l$, the particle and the hole are confined, forming a dipole condensate.  In the opposite limit, where the correlation length $l$ becomes divergent, the particle and the hole become deconfined. This is accompanied by the rise of  the single-particle condensate and the single-particle order parameter $\langle b_{\bf i}\rangle$ becomes finite.  

The prevalence of a dipole condensate can also be understood from a different perspective. The order parameter of a dipole condensate can be regarded as the Fourier transform of the momentum distribution,  
\begin{equation}
    \langle D_{{\bf i},{\bf s}}\rangle
    =\frac{1}{L}\sum_{\bf k}e^{i \bf k\cdot(R_{\bf i+\bf s}-R_{\bf i})}n_{\bf k},\label{dnk}
\end{equation}
where $n_{\bf k}=\langle a^\dagger_{\bf k}a_{\bf k}\rangle$ is the momentum distribution, $a^\dagger_{\bf k}$($a_{\bf k}$) is the creation (annihilation) operator in the momentum space, $L$ is the number of lattice sites, and ${\bf R}_{{\bf i}+{\bf s}}$ is the coordinate of site ${\bf i}+{\bf s}$. We have considered homogeneous systems such that $\langle a^\dagger_{\bf k} a_{\bf q}\rangle=0$ if ${\bf k}\neq {\bf q}$. If a single-particle condensate exists, $n_{\bf k}\sim N_0\delta_{\bf k}$, where $N_0$ is of the order of the total particle number $N$. The above equation immediately tells us that $\langle b^\dagger_{\bf i} b_{\bf{i}+\bf {s}}\rangle$ is of the order of $n_0=N/L$, signifying a dipole condensate. This is what one naturally expects, as a finite single-particle order parameter $\langle b_{\bf i}\rangle$ automatically leads to a finite two-body order parameter $\langle b^\dagger_{\bf i} b_{\bf j}\rangle\sim \langle b^\dagger_{\bf i}\rangle \langle b_{\bf j}\rangle$. But a single-particle condensate is only a sufficient,  not a necessary, condition for a dipole condensate. Even when the single-particle condensate vanishes and $n_{\bf k}$ may have a broad distribution in the momentum space, the right-hand side of Eq.~(\ref{dnk}) may still lead to a constructive interference such that $\langle b^\dagger_{\bf i} b_{\bf{i}+\bf {s}}\rangle$ is of the order of $n_0$. For instance, $n_{\bf k}$ is a constant in a ``Fermi" sea. We will discuss such examples in section ({\bf III.A}). 


We emphasize that the existence of a dipole condensate does not require the dipole conservation law in the system. If the dipole is conserved, for instance, in the DBHMs, the charge is strictly immobile, as the movement of a charge inevitably creates an extra dipole.  Even in the absence of the dipole conservation law, the motion of a charge may still be constrained. In a generic normal phase of bosons, either thermal or quantum fluctuations destroy the off-diagonal long-range order, and the correlation length $l$ becomes finite.  A dipole condensate could therefore be defined.

\subsection{Examples of dipole condensates} 
\begin{figure}[htbp]
\centering  
\includegraphics[width=.95\columnwidth]{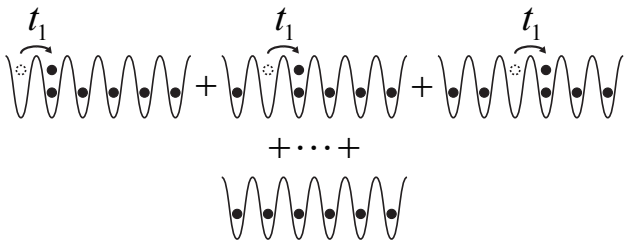}
\caption{An illustration of the ground state 
of BHM in the limit $t_1/U\ll1$, which is a superposition of Mott insulator and particle-hole pair excitations.}
\label{ansatz}
\end{figure}

We first consider a conventional Bose-Hubbard model(BHM),
\begin{equation}
    H_1=-\sum_{\langle {\bf i},{\bf j}\rangle }(t_1 b_{\bf i}^\dagger b_{{\bf j}}+h.c.)+\frac{U}{2}\sum_{\bf i} n_{\bf i}(n_{\bf i}-1),\label{bhm}
\end{equation}
where $\langle {\bf i},{\bf j}\rangle $ denotes the nearest neighbor sites. At zero temperature, when $t_1/U$ is larger than a critical value $(t_1/U)_c$, the ground state is a condensate, $\langle b_{\bf i}\rangle\neq 0$. When $t_1/U<(t_1/U)_c$, the off-diagonal long-range order and the single-particle condensate vanish. The ground state becomes a Mott insulator~\cite{Fisher1989, Greiner2002}. Nevertheless, short-range correlations exist provided that $t_1/U$ is finite. For instance, when
$t_1/U\ll1$, 
the ground state can be well approximated by 
\begin{equation}
    |\Psi_1\rangle\approx|\text{MI}\rangle+\frac{t_1}{U}\sum_{\langle {\bf i},{\bf j}\rangle}(\hat{b}_{\bf i}^\dagger\hat{b}_{\bf j}+h.c.)|{\text{MI}}\rangle,\label{Short}
\end{equation}as shown in Fig.~\ref{ansatz}.

The second term in Eq.~(\ref{Short}) leads to short-range correlations even deep in the Mott insulator regime, which has been observed in experiments~\cite{Bloch2005}. Using Eq.~(\ref{Short}), it is straightforward to obtain
\begin{equation}
    \langle D_{{\bf i},\hat{\bf x}}\rangle=\langle D_{{\bf i},\hat{\bf y}}\rangle=\frac{t_1}{U}2n_0(n_0+1),
\end{equation}
from which we conclude that the Mott insulator with short-range correlation is readily a dipole condensate. The reduced one-body density matrix of the dipole, $\langle D^\dagger_{{\bf i},{\bf s}}D_{{\bf j},{\bf r}}\rangle$, approaches a constant in the large $|\bf i-\bf j|$ limit. Here, ${\bf s}$ and ${\bf r}$ denote $\hat{\bf x}$ or $\hat{\bf y}$. 

More generically,  we could evaluate the reduced single-particle density matrix $\rho_1$,
\begin{equation}
\rho_1(\bf i,\bf j)=\begin{cases}
    n_0, &|{\bf i}-{\bf j}|= 0,\\
\frac{t_1}{U}2n_0(n_0+1), &|{\bf i}-{\bf j}|=1.\\
\end{cases}
\end{equation}
The momentum distribution is simply the Fourier transform of  $\rho_1$,
\begin{equation}
    n({\bf k})= n_0+4\frac{t_1}{U}n_0(n_0+1)(\cos{k_x d}+\cos{k_y d}),\label{nkMI}
\end{equation}
where $d$ is the lattice constant.
Substituting Eq.~(\ref{nkMI}) to Eq.~(\ref{dnk}), we see that the first term, $n_0$, does not contribute to the right hand side of Eq.~(\ref{dnk}), as $\sum_{\bf k}e^{i \bf k\cdot(R_{\bf i+\bf s}-R_{\bf i})} =0$ for any finite ${\bf s}$. In contrast, the second term on the right-hand side of Eq.~(\ref{nkMI}) leads to a constructive interference on the right-hand side of Eq.~(\ref{dnk}). As such, $\langle D_{{\bf i},\hat{\bf x}}\rangle$ and $\langle D_{{\bf i},\hat{\bf y}}\rangle$ become finite despite that the single particle condensate is absent. 

When $t_1/U$ increases, Eq.~(\ref{nkMI}) is no longer a good approximation. Nevertheless, the same conclusion about the dipole condensate applies. The only quantitative difference is that $\langle D_{{\bf i},{\bf s}}\rangle$ becomes finite for a generic $\bf{s}$. The width of the relative wavefunction of the dipole increases while the dipole condensate remains finite. To be more explicit, the reduced one-body density matrix of the dipole, $\langle D^\dagger_{{\bf i},{\bf s}}D_{{\bf j},{\bf r}}\rangle$, 
remains finite when $|{\bf i}-{\bf j}|\rightarrow \infty$ for generic ${\bf s}$ and ${\bf r}$.  When $t_1/U=(t_1/U)_c$, dipoles become deconfined and a single-particle condensate arises. In the opposite limit, $t_1/U=0$ and the dipole condensate vanishes. 

Another example is the normal phase at finite temperatures. For instance, considering the high-temperature regime, $T\gg t_1\gg U$, the momentum distribution can be written as 
\begin{equation}
   n_{\bf k}=e^{\beta (\mu-\epsilon_{\bf k})} ,\label{nkft}
\end{equation}
where $\mu$ is the chemical potential and $\epsilon_{\bf k}$ is the single-particle energy. In the dilute limit, $\epsilon_{\bf k}=\hbar^2k^2/(2m^*)$, where $k$ is the magnitude of ${\bf k}$ and $m^*$ is the effective mass at the band bottom. Substituting Eq.~(\ref{nkft}) to Eq.~(\ref{dnk}), we obtain
\begin{equation}
\langle D_{{\bf i},{\bf s}}\rangle=n_0\frac{\sum_{\bf k}e^{-i{\bf k}\cdot({\bf R}_{\bf i+\bf s}-{\bf R}_{\bf i})}e^{-\beta \epsilon_{\bf k}}}{\sum_{\bf k}e^{-\beta \epsilon_{\bf k}}}.\label{Dis}
\end{equation}
The numerator in Eq.~(\ref{Dis}) leads to a constructive interference such that $\langle D_{{\bf i},{\bf s}}\rangle$ become finite for any finite $\bf s$. For instance, we may consider a one-dimensional gas at the high temperature regime $T\gg T_{1D}$, where 
$T_{1D}\equiv 2\pi{\hbar^2{n_0}^2}/{m^*d^2}$.
It is straightforward to obtain the momentum distribution $n_k= {\sqrt{{T_{1D}}/{T}}}e^{-\epsilon_k/T}$, and therefore $\langle D_{{\bf i},\hat{\bf x}}\rangle=n_0e^{-{m^*Td^2}/{2\hbar^2}}$. We thus conclude that the normal phase at finite temperatures is readily a dipole condensate as well. Only in the limit of the infinite temperature, $\langle D_{{\bf i},\hat{\bf x}}\rangle$ is suppressed down to zero and the dipole condensate vanishes.

\subsection{Self-proximity effects}

The above results of dipole condensate can be understood as a self-proximity effect. The proximity effect has played a vital role in the study of superconductivity~\cite{McMillan1968}. When a superconductor is placed in contact with a 
normal metal, Cooper pairs tunnel through the interface. Such a tunneling process acts as a linear term of the creation or  annihilation of Cooper pairs and thus automatically induces superconductivity in the normal state. 
Here, we can think about BHM in a similar means. The first term in Eq.~(\ref{bhm}), i.e., the kinetic energy of single particles, is readily a linear term of the creation (or annihilation) operator of dipoles. In the mean-field approach, it corresponds to a linear term of the order parameter of the dipole. As such, no matter how weak it is, a finite dipole order parameter must be produced. 

\begin{figure}[htbp]
\centering  
\includegraphics[width=.8\columnwidth]{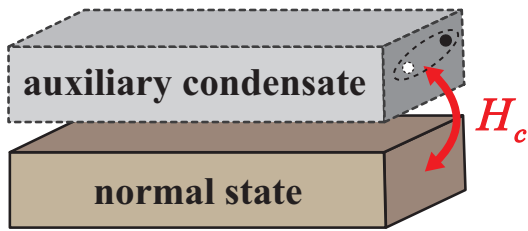}
\caption{A sketch of the self-proximity effect. A normal state  (the lower layer) 
is coupled to an auxiliary condensate (the upper layer) by the tunnelings of particle-hole pairs that automatically induce dipole condensation in the normal state. Here, no external condensate is required in reality and the single-particle kinetic energy plays the role of the auxiliary condensate.}
\label{proximity}
\end{figure}

To be more explicit, we may imagine that there is an auxiliary condensate coupled to the system of interest, as shown by Fig.~\ref{proximity}. The Hamiltonian of the composite system is written as $H=\tilde{H}_1+H_a+H_c$, where 

\begin{equation}
\tilde{H}_1=\frac{U}{2}\sum_{\bf i} n_{\bf i}(n_{\bf i}-1),
\end{equation}
\begin{equation}
    H_a=-\sum_{\langle \bf i,j\rangle}(t_a c^\dagger_{\bf i}c_{\bf j}+h.c.),
\end{equation}
and
\begin{equation}
    H_c=- \sum_{\langle{\bf i, j}\rangle}(\Omega_{\bf{ i,j}} b^\dagger_{\bf i} b_{\bf j} c^\dagger_{\bf j}c_{\bf i}+h.c.).
\end{equation}
Because of the absence of kinetic energy, the ground state of $\tilde{H}_1$ is a Mott insulator with no single-particle condensation or dipole condensation. 
At finite temperatures, both condensates are still absent. 
In contrast, the lack of interactions in $H_a$ tells us that the ground state of the auxiliary system is a condensate $\langle c_{\bf i}\rangle=\phi$. $H_c$ denotes the hopping of particle-hole pairs between the bosonic system of interest and the auxiliary condensate, as analogous to the particle-particle pair coupling in conventional proximity effects of superconductors. The Mott insulator and the auxiliary condensate play the roles of the normal metal and the superconductor in the ordinary proximity effect.

The condensation of the auxiliary system allows us to replace $c^\dagger_{\bf j} c_{\bf i}$ by its 
expectation value and $H_c$ becomes 
\begin{equation}
    H_{c,\text{mf}}=- \sum_{\langle\bf{i,j}\rangle}(\Omega_{\bf{i,j}}|\phi|^2 b^\dagger_{\bf i} b_{\bf j} +h.c.).
\end{equation}
This is precisely a counterpart of the induced Cooper pairing in the conventional proximity of superconductors, where $|\phi|^2$ plays the role of the order parameter of the superconductor. As such, an auxiliary condensate automatically induces a dipole condensate, $\langle b^\dagger_{\bf i}b_{\bf j}\rangle \neq 0$, regardless of how strong the finite interaction $U$ is. The same conclusion applies to finite temperatures where the Mott insulator is replaced by a generic normal state.

As a comparison, we consider a different coupling between the auxiliary condensate and the system of interest,
\begin{equation}
    H'_c=- \sum_{ {\bf i}}(\Omega'_{\bf i} b^\dagger_{\bf i} c_{\bf i}+h.c.).
\end{equation}
Now the Mott insulator and the auxiliary condensate are coupled by the single-particle tunneling. A finite $\langle c_{\bf i}\rangle=\phi'$ gives rise to the mean-field Hamiltonian, 
\begin{equation}
    H'_{c,\text{mf}}=- \sum_{ {\bf i}}(\Omega'_{\bf i}\phi' b^\dagger_{\bf i} +h.c.).
\end{equation}
$\Omega'_{\bf i}\phi'$ and its hermitian conjugate serve as the source and the drain of single particles, respectively. A finite $\phi'$ thus produces a finite  $\langle b^\dagger_{\bf i}\rangle$, regardless of how large $U$ is. This time, a single-particle condensate is induced. We define $H_0=H'_{c,\text{mf}}+\tilde{H}_1$, which can be written as
\begin{equation}
    H_0=-\sum_i (t_0 b^\dagger_{\bf i}+h.c.)+\frac{U}{2}\sum_in_{\bf i}(n_{\bf i}-1).
\end{equation}
We have considered a site-independent $\Omega_{\bf i}'\equiv\Omega'$ to simplify notations, and $t_0=\Omega'\phi'$. The ground state of $H_0$ is always a single-particle condensate even in the large $U$ limit, as analogous to the dipole condensate as the ground state of $H_1$ in the large $U$ limit.

Whereas the above discussions resemble that of the conventional proximity effect,  we must emphasize an intrinsic difference.  Here, no other quantum systems are required in reality and the auxiliary condensate is introduced only for the purpose of understanding the underlying physics. The proximity effect is a self-induced one. The kinetic energy of single particles in $H_1$, i.e., $b^\dagger_{\bf{i}}b_{\bf j}$, readily provides the source of the dipole condensate, similar to $b^\dagger_{\bf i}$ as the single-particle source in $H_0$. We thus name this effect a self-proximity effect. Such an effect also exists in a more generic situation when we consider an arbitrary multipolar condensate. This will be discussed in Section {\bf V}. 

We also would like to emphasize that the phase of a dipole condensate induced by the self-proximity effect is fixed by the single-particle tunneling, unlike  the ground state of DBHM where the spontaneous symmetry breaking allows an arbitrary phase of the dipole condensate. Controlling the phase of the single-particle tunneling thus provides experimentalists with a unique means to tune and twist the phase of a dipole condensate. Turning on correlated hoppings then gives rise to a dipolar Josephson effect. This will be discussed in section ({\bf IV}).

\section{Non-equilibrium preparation of dipole condensates} 

\subsection{Bosonic metal}

\begin{figure}[htbp]
\centering  
\includegraphics[width=.9\columnwidth]{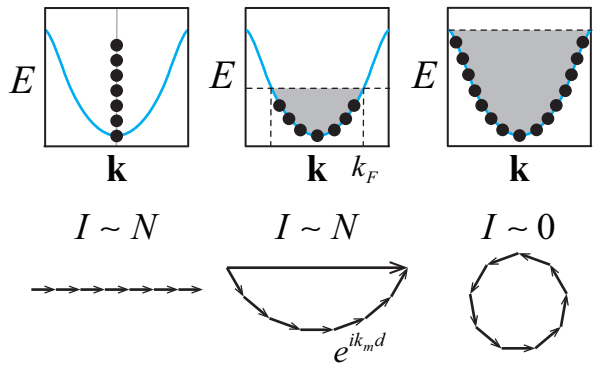}
\caption{Interference patterns determining $I\equiv\sum_{\bf k}e^{i \bf k\cdot(R_{\bf i+\bf s}-R_{\bf i})}n_{\bf k}$ for different momentum distributions $n_{\bf k}$. These three panels denote the results of a single-particle condensate (left), a bosonic metal (middle), and a band insulator (right).}
\label{interfere}
\end{figure}

Eq.~(\ref{dnk}) tells us that a broad range of $n_{\bf k}$ may produce a dipole condensate. In particular, we could consider $n_{\bf k}$ unattainable at equilibrium. In fact, non-equilibrium dynamics have been widely used as an efficient tool to prepare novel quantum states in laboratories. Here, we consider a bosonic metal, whose momentum distribution resembles that of non-interacting fermions,
\begin{equation}
n_{\bf k}=
\begin{cases}
    n_0,& \epsilon_{\bf k}<E_F,\\
    0, &\epsilon_{\bf k}>E_F,
\end{cases}
\end{equation}
where $E_F$ is the ``Fermi" energy. We have previously discussed that using the momentum distribution of a single particle condensate, $n_{\bf k}\sim N_0 \delta_{\bf k}$, $I\equiv\sum_{\bf k}e^{i \bf k\cdot(R_{\bf i+\bf s}-R_{\bf i})}n_{\bf k}$ is of the order of $N$ and thus leads to a finite order parameter of dipole. Here, using $n_{\bf k}$ of a bosonic metal, $I$ is also of the order of $N$ and $\langle D_{{\bf i},{\bf s}}\rangle $ is of the order of $N/L$. A pictorial description of such constructive interference is shown in Fig.~\ref{interfere}. A bosonic metal is thus a dipole condensate.

It is worth mentioning that, to deliver a dipole condensate, the occupied states do not need to be the ones with the lowest single-particle energies. For instance, the bosonic cloud can be shifted by a constant ${\bf k}_0$ in the momentum space. The dipole order parameter becomes a function of ${\bf k}_0$, 
\begin{equation}
    \langle D_{{\bf i},{\bf s}}\rangle({\bf k_0})=e^{i \bf k_0\cdot(R_{\bf i+\bf s}-R_{\bf i})}\langle D_{{\bf i},{\bf s}}\rangle(0).
\end{equation}
In other words, the dipole condensate acquires a ${\bf s}$-dependent phase. $n_{\bf k}$ could also be a more generic function in addition to a step function. The previously discussed finite temperature result at thermal equilibrium is such an example.

When $E_F$ continues to increase, all states in the Brillouin zone (BZ) are eventually occupied, and the bosonic metal turns into a bosonic band insulator,
\begin{equation}
    |{\text{BI}}\rangle=\prod_{{\bf k}\in BZ }\frac{a^{\dagger n_0}_{\bf k}}{\sqrt{n_0!}}|0\rangle.\label{bi}
\end{equation}
Previous studies have considered such a state in one dimension with $n_0=1$ in the context of a $N$-port Hong-Ou-Mandel interferometer in quantum optics~\cite{Lim2005, Buchleitner2010}. It has been shown that certain output events are strictly prevented. 
In our language, 
this corresponds to dipole conservation. Whereas the real-space representation of $|{\text{BI}}\rangle$ includes many product states, the dipole moment mod $N$ is conserved, i.e., ($P$ \text{mod} $N$)=0, where $P=\sum_i i n_i$ is the dipole moment of a product state with the occupation number at site $i$ denoted by $n_i$. Any state that does not satisfy this conservation law is, therefore, suppressed in the output of the interferometer. Here, Eq.~(\ref{dnk}) shows that $\langle D_{{\bf r},{\bf s}}\rangle$ becomes zero for any finite ${\bf s}$ and the dipole condensate vanishes. Adding a finite doping creates a bosonic metal and thus a dipole condensate. 

\subsection{Multiport Hong-Ou-Mandel interferometers}
Now we discuss how to create a bosonic metal in laboratories. A recent experiment has implemented an adiabatic process to transfer the occupancy in the real space to the momentum space~\cite{Saxberg2022}. This method could, in principle, deliver a bosonic metal by populating some lattice sites in the real space and then adiabatically converting them to desired momentum states. However, at the final stage, ${\bf k}$ and $-{\bf k}$ are typically degenerate, imposing a challenge to retain adiabaticity. Moreover, it is time-consuming to adopt adiabatic processes in reality and a shortcut to adiabaticity is often appreciated. To this end, we present a preparation method using non-equilibrium quantum dynamics. The idea is to design a  Hamiltonian as the generator of the required transform matrix. 
On a 1D lattice with $L$ sites, the required transform matrix is the Discrete Fourier Transform (DFT) matrix,
\begin{equation}
   \mathcal{F}   = \frac{1}{\sqrt{L}}\left(\begin{array}{ccccc}
       1  &  1  &   \cdots   &   1   &   1 \\
       1  &  \omega  &  \cdots & \omega^{L-2} & \omega^{L-1}\\
       \vdots & \vdots & \cdots & \vdots & \vdots\\
       1  & \omega^{L-1} & \cdots & \omega^{(L-2)(L-1)} & \omega^{(L-1)(L-1)}\\
    \end{array}\right)\label{F}
\end{equation}
where $\omega=e^{2\pi i/L}=\cos(2\pi/L)+i\sin(2\pi/L)$. Acting $\mathcal{F}$ on creation operators in  the real space 
provides operators in the momentum space, namely
\begin{equation}
   \mathcal{F} \left(\begin{array}{c}
       b_1^\dagger   \\
       b_2^\dagger   \\
       \vdots \\
       b_L^\dagger  \\
    \end{array}\right) = \left(\begin{array}{c}
       a_{k_1}^\dagger   \\
       a_{k_2}^\dagger   \\
       \vdots \\
       a_{k_L}^\dagger  \\
    \end{array}\right),
\end{equation}
where ${k_i}=2\pi(i-1)/L$ and $i=1,2,...,L$.
If all lattice sites are initially occupied by one particle, after applying $\mathcal{F}$, we obtain a band insulator. Similarly, this method could deliver an arbitrary distribution of particles in the momentum space. For instance, to obtain a bosonic metal, experimentalists just need to fill the lattice sites corresponding to the momentum states within the ``Fermi" sea. 

To realize a DFT in experiments, we implement an interesting property of the DFT matrix. 
If we define
\begin{equation}
    \mathcal{H}=\frac{1}{2}(1-(1+i)\mathcal{F}+\mathcal{F}^2-(1-i)\mathcal{F}^3),
\end{equation} 
it has been shown that~\cite{Aristidou2007}
\begin{equation}
    \mathcal{F}=e^{i\mathcal{H}\pi/2}\label{pulse}.
\end{equation}
If we regard $\mathcal{H}$ as a generator of time translation and define a Hamiltonian
\begin{equation}
    H=-\left(\begin{array}{cccc}
     b_1^\dagger  &  b_2^\dagger &  \ldots  &  b_L^\dagger\\
    \end{array}\right)\mathcal{H}\left(\begin{array}{c}
         b_1 \\
         b_2 \\
         \vdots\\
         b_L \\
    \end{array}\right),\label{H}
\end{equation}
$\mathcal{F}$ is a $\pi/2$ pulse that establishes a one-to-one mapping between lattice sites and the momentum states via $k_j=2\pi(j-1)/L$, according to $e^{-iH\pi/2}b_j^\dag e^{iH\pi/2}=a_{k_j}^\dag$.
If the initial state is a Mott insulator, we obtain
\begin{equation}
    e^{-iH\pi/2}|\text{MI}\rangle=|\text{BI}\rangle.
\end{equation}
Alternatively, if the lattice sites are partially filled in the initial state, 
$|\text{PF}\rangle=\prod_{i=1}^{i_{M}} {b_i^{\dagger n_0}}/{\sqrt{n_0!}} |0\rangle$, where $i_M<L$.  $b^{\dagger}_i$ ($i=1,2,...i_{M})$ is mapped to $a^\dagger_{k_i}$ by this $\pi/2$ pulse, where $|k_i|<k_F$ and $k_F$ is the ``Fermi" momentum. $|\text{PI}\rangle$ becomes the desired bosonic metal,
\begin{equation}
    e^{-iH\pi/2}|\text{PI}\rangle=|\text{BM}\rangle\equiv \prod_{k<k_F}\frac{a^{\dagger n_0}_{k}}{\sqrt{n_0!}}|0\rangle.
\end{equation}

As an example, when $L=2$, the DFT matrix is simply
\begin{equation}
    \mathcal{F} =\frac{1}{\sqrt{2}}\left(\begin{array}{cc}
       1  &  1\\
       1  & -1
    \end{array}\right).
\end{equation} 
Based on Eq.(\ref{H}), the Hamiltonian needs to be
\begin{equation}
    H=-\left(\begin{array}{cc}
     b_1^\dagger  &  b_2^\dagger\\
    \end{array}\right)\left(\begin{array}{cc}
       1-\frac{1}{\sqrt{2}}  &  -\frac{1}{\sqrt{2}}\\[6pt]
       -\frac{1}{\sqrt{2}} & 1+\frac{1}{\sqrt{2}}
    \end{array}\right)\left(\begin{array}{c}
         b_1 \\
         b_2 \\
    \end{array}\right).
\end{equation} It is straightforward to verify that $e^{-iH\pi/2}b_1^\dagger e^{iH\pi/2}=(b_1^\dagger+b_2^\dagger)/\sqrt{2}$ and $e^{-iH\pi/2}b_2^\dagger e^{iH\pi/2}=(b_1^\dagger-b_2^\dagger)/\sqrt{2}$. Each of these two lattice sites is uniquely mapped to a state in the momentum space.  

\begin{figure}[htbp]
    \centering
    \includegraphics[width=0.8\columnwidth]{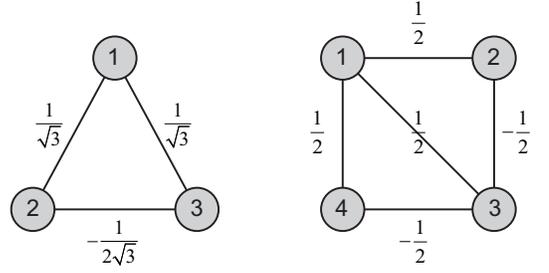}
    \caption{Demonstrations of the required Hamiltonians for the non-equilibrium preparation of a bosonic metal in a lattice with $L=3$ and $L=4$ sites. The numbers on bonds denote the tunneling strengths between lattice sites.}
    \label{sitecoupling}
\end{figure}

The above discussion can be applied to an arbitrary $L$. Here we show results of Eq.~(\ref{H}) when $L=3$,
 \begin{equation}
    H=-\left(\begin{array}{ccc}
     b_1^\dagger  &  b_2^\dagger  &  b_3^\dagger  
    \end{array}\right)\left(\begin{array}{ccc}
       1-\frac{1}{\sqrt{3}}  &  -\frac{1}{\sqrt{3}} &  -\frac{1}{\sqrt{3}}  \\[6pt]
       -\frac{1}{\sqrt{3}} &  1+\frac{1}{2\sqrt{3}}  &  \frac{1}{2\sqrt{3}}  \\[6pt]
       -\frac{1}{\sqrt{3}}  &  \frac{1}{2\sqrt{3}}  &  1+\frac{1}{2\sqrt{3}}  \\[6pt]
    \end{array}\right)\left(\begin{array}{c}
         b_1 \\
         b_2 \\
         b_3
    \end{array}\right)
\end{equation}, and when $L=4$,

\begin{equation}
    H=-\left(\begin{array}{cccc}
     b_1^\dagger  &  b_2^\dagger  &  b_3^\dagger  &  b_4^\dagger\\
    \end{array}\right)\left(\begin{array}{cccc}
       \frac{1}{2}  &  -\frac{1}{2} &  -\frac{1}{2}  &  -\frac{1}{2}\\[6pt]
       -\frac{1}{2} &  1  &  \frac{1}{2}  &  0\\[6pt]
       -\frac{1}{2}  &  \frac{1}{2}  &  \frac{1}{2}  &  \frac{1}{2}\\[6pt]
       -\frac{1}{2}  &  0  &  \frac{1}{2}  &  1\\
    \end{array}\right)\left(\begin{array}{c}
         b_1 \\
         b_2 \\
         b_3 \\
         b_4
    \end{array}\right).
\end{equation}
The off-diagonal elements of $H$ represent the tunneling strengths among lattice sites, as shown in Fig.~\ref{sitecoupling}. If these lattice sites are aligned in a one-dimensional chain, tailored long-range tunnelings are required. Such long-range tunnelings can be created using nanophotonics, cavities, trapped ions, or superconducting circuits ~\cite{Hung2016, Baumann2010, Islam2013, Xu2020, Monika2021}. Alternatively, movable optical tweezers could be implemented to design tunnelings between any arbitrary pair of lattice sites in the same manner as creating the graph states ~\cite{Bluvstein2022}. 

This method can be generalized to  higher dimensions. For example, on a $L_x\cross L_y$ square lattice, $\mathcal{F}$ is defined as
\begin{equation}
   \mathcal{F} \left(\begin{array}{c}
       b_{1,1}^\dagger   \\
       b_{1,2}^\dagger   \\
       \vdots \\
       b_{i,j}^\dagger    \\
       \vdots \\
    \end{array}\right) = \left(\begin{array}{c}
       a_{k_1,k_1}^\dagger   \\
       a_{k_1,k_2}^\dagger   \\
       \vdots \\
       a_{k_i,k_j}^\dagger  \\
       \vdots \\
    \end{array}\right),
\end{equation}where ${k_{i(j)}}=2\pi(i(j)-1)/L_{y(x)}$, and $i(j)=1,2,...,L_{y(x)}$. Since $\mathcal{F}$ describes 2D Fourier transformation, it can be written as a product of two 1D Fourier transformations, namely $\mathcal{F}=\mathcal{F}_y\mathcal{F}_x$. $\mathcal{F}_x$ acts on chains along the $x$ direction, and it is a block diagonal matrix including $L_y$ blocks,
\begin{equation}
   \mathcal{F}_x   = \left(\begin{array}{ccccc}
       \mathcal{D}_x  &    &      &   &  \\
               & \ddots &  & & \\
         &   & \mathcal{D}_x  & & \\
        &  &  & \ddots & \\
         &  &  & &\mathcal{D}_x \\
    \end{array}\right),
\end{equation}where $\mathcal{D}_x$ is a $L_x\cross L_x$ block acting on each chain in the $x$ direction, as shown by Eq.~(\ref{F}) with $L$ replaced by $L_x$. Similarly, $\mathcal{F}_y$ acts on chains along $y$ direction. It can be block diagonalized by rearranging the order of the basis $b_{i,j}^\dagger$ based on the index $j$. Both $\mathcal{F}_x$ and $\mathcal{F}_y$ can be realized using $\pi/2$ pulse, as we previously discussed. As an example, we consider the $L_x=L_y=2$ case,
\begin{equation}
\begin{aligned}
        \mathcal{F} =&\frac{1}{2}\left(\begin{array}{cccc}
       1  &  1  &  1  &  1\\
       1  &  -1  &  1  &  -1\\
       1  &  1  &  -1  &  -1\\
       1  &  -1  &  -1  &  1\\
    \end{array}\right)\\
    =&\frac{1}{\sqrt{2}}\left(\begin{array}{cccc}
       1  &  0  &  1  &  0\\
       0  &  1  &  0  &  1\\
       1  &  0  &  -1  &  0\\
       0  &  1  &  0  &  -1\\
    \end{array}\right)\cdot\frac{1}{\sqrt{2}}\left(\begin{array}{cccc}
       1  &  1  &  0  &  0\\
       1  &  -1  &  0  &  0\\
       0  &  0  &  1  &  1\\
       0  &  0  &  1  &  -1\\
    \end{array}\right)\\
    \equiv&\mathcal{F}_y\mathcal{F}_x.
\end{aligned}
\end{equation}
As $\mathcal{F}_x$ and $\mathcal{F}_y$ commute with each other, we may apply these two $\pi/2$ pulses in an arbitrary order. The required Hamiltonian for each step can be obtained following the previous discussions about one dimension.

\section{Dipolar Josephson effect}
To access macroscopic quantum phenomena, it is crucial to manipulate the phase of a condensate. Here, our results can be utilized to imprint tailored phase patterns to dipole condensates. In BHM as shown in Eq.~(\ref{bhm}), an additional phase can be added to the tunneling $t_1$, $t_1\rightarrow t_1 e^{i\theta}$, using synthetic gauge fields~\cite{Jaksch2003, Dalibard2011, Goldman2014}. This amounts to providing the dipole condensate with a phase, $\langle D_{{\bf i},{\bf i}+\hat{\bf x}}\rangle\rightarrow \langle D_{{\bf i},{\bf i}+\hat{\bf x}}\rangle e^{i\theta}$. This phase can be further made position dependent. For instance, we consider two chains, each of which is described by a BHM, and the tunnelings $t_1$ have different phases. This phase difference could be used to create supercurrents of dipoles and dipolar Josephson effect.

\begin{figure}[htbp]
\centering  
\includegraphics[width=0.95\columnwidth]{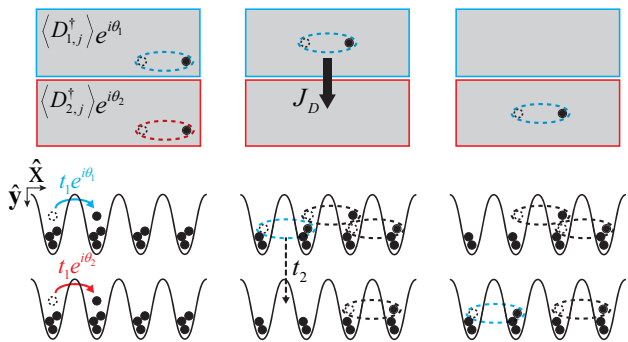}
\caption{A proposed experimental scheme to show the dipolar Josephson effect of planons. In step 1(left), the system is initialized at the ground state of BHM in each chain. 
Different phases $\theta_1\neq\theta_2$ are imprinted to these two chains. 
In step 2(middle), 
the ring-exchange interaction is turned on at $\tau=0$ and allows planons to tunnel between these two chains. In step 3(right), a supercurrent of dipoles between two chains is measured by tracing the dipole moment of each chain as time goes by.}
\label{2chain}
\end{figure}

The dipolar Josephson effect is shown in Fig.~\ref{2chain}. Initially, the Hamiltonian is written as 
\begin{equation}
    H_i=-\sum_{l,j}(t_1e^{i\theta_l}b^\dagger_{l,j+1}b_{l,j}+h.c.)+\frac{U}{2}\sum_{l,j}n_{l,j}(n_{l,j}-1),\label{2chainHi}
\end{equation}where $l=1, 2$ denote the upper and the lower chain, respectively.
As we previously discussed, when $t_1/U<(t_1/U)_c$, the ground state of $H_i$ is a dipole condensate. Here, a critical ingredient is that a phase difference $\Delta\theta\equiv\theta_1-\theta_2$ between the upper chain and the lower chain has been imprinted to the system. 
At time $\tau=0$, we turn on a ring exchange interaction and change the Hamiltonian to the DBHM,
\begin{equation}
\begin{aligned}
    H_f=&-\sum_j(t_2{b_{1,j}^\dagger b_{2,j+1}^\dagger b_{1,j+1} b_{2,j}+h.c.})\\
    &+\frac{U}{2}\sum_{l,j}n_{l,j}(n_{l,j}-1).\label{2chainHf}
\end{aligned}
\end{equation}The ground state of $H_i$ is no longer an eigenstate of $H_f$, the dipole tunneling $t_2$ and a finite $\Delta\theta$ jointly create a dipolar Josephson effect. 

We define the dipole moment of each chain, 
\begin{equation}
    P_1(\tau)=\sum_j jn_{1,j},\,\,\,\,\, P_2(\tau)=\sum_j jn_{2,j}.
\end{equation}
And the total dipole moment is
\begin{equation}
    P(\tau)=\sum_{l,j} j n_{l,j}(\tau).
\end{equation}

\begin{figure}[htbp]
\centering  
\subfigure[]{
\label{dm2}
\includegraphics[width=0.48\columnwidth]{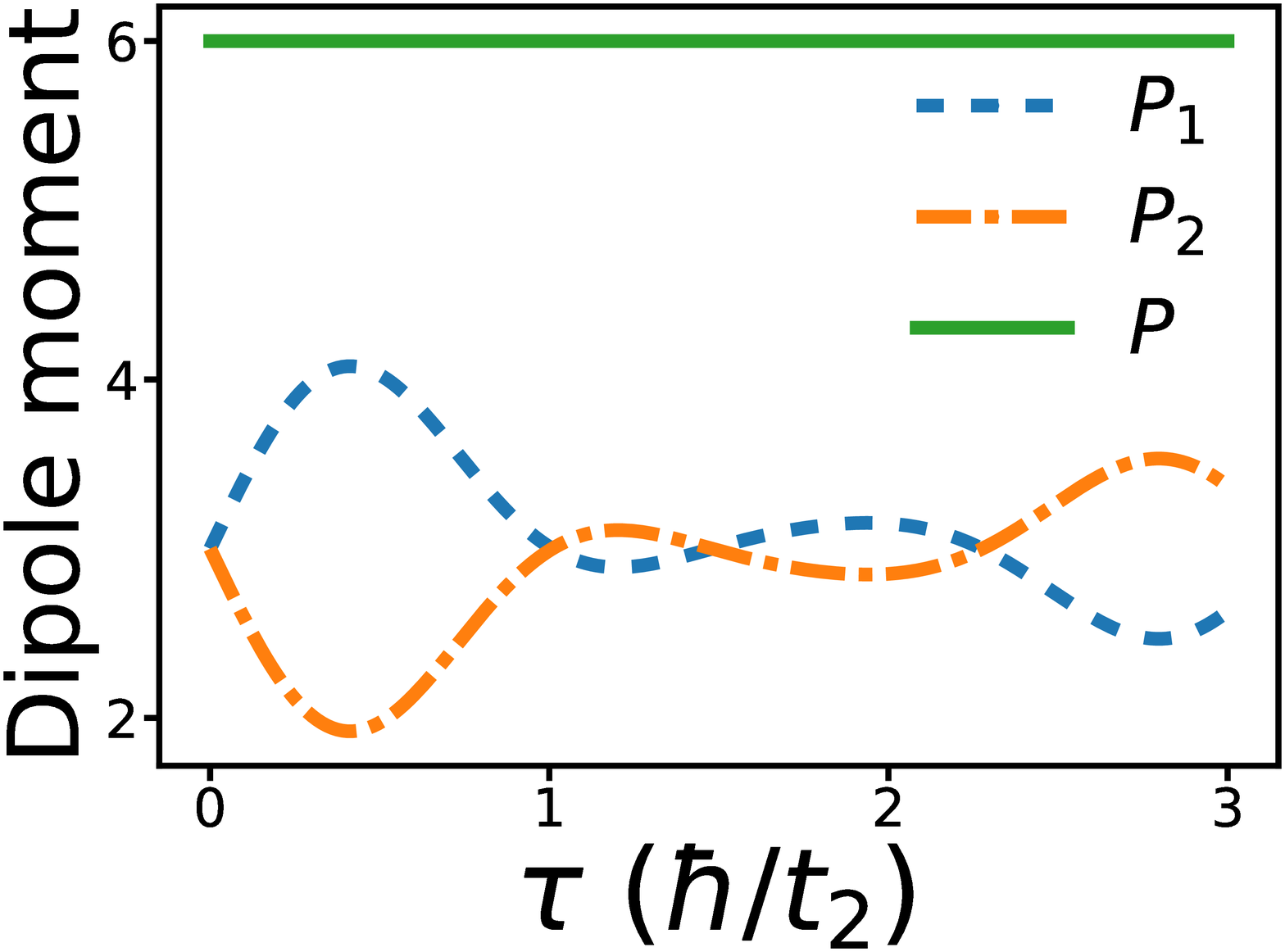}}
\hspace{-5pt}
\subfigure[]{
\label{dc2}
\includegraphics[width=0.48\columnwidth]{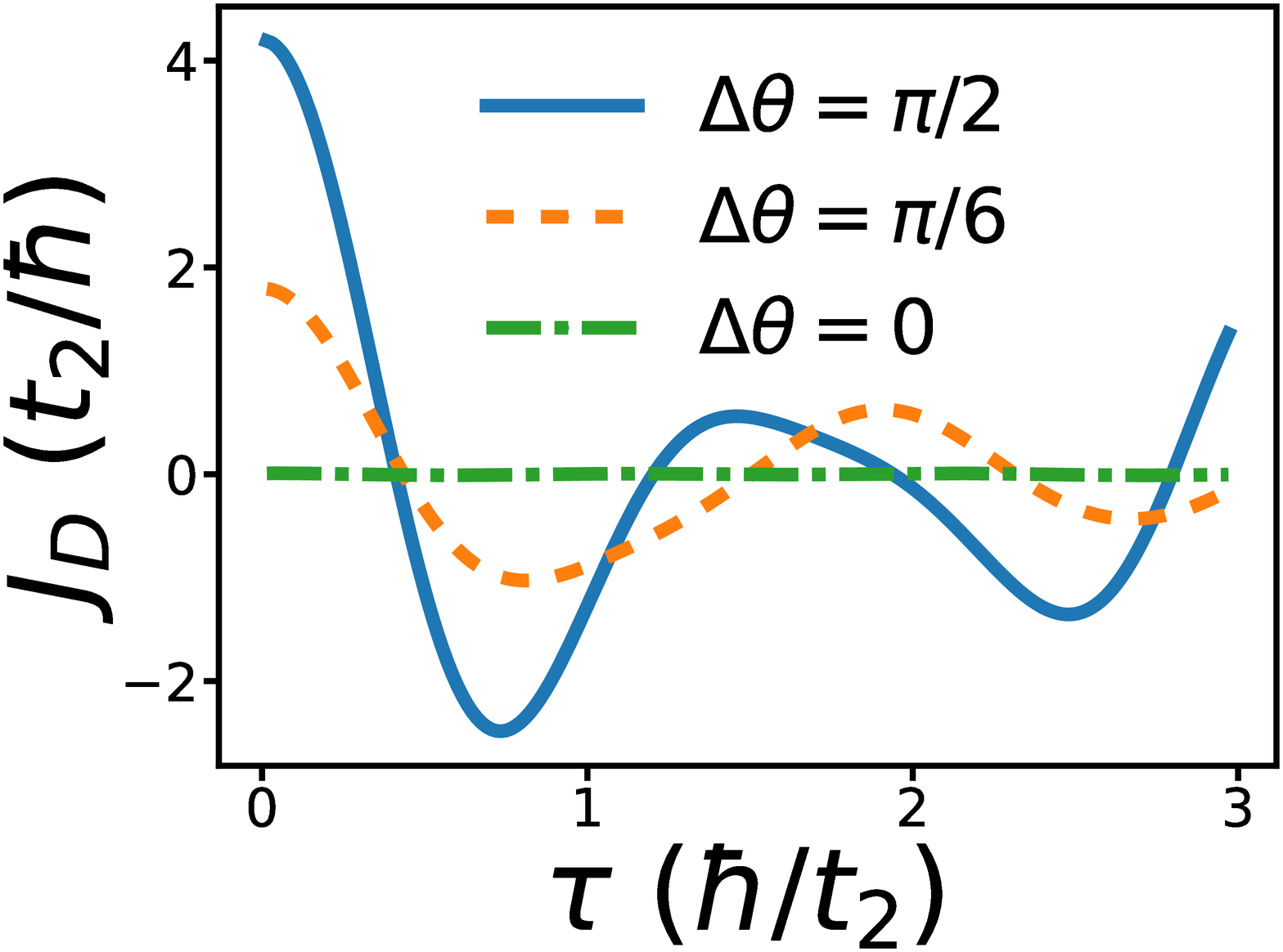}}
\caption{(a) Numerical results of the dipole moments of both chains $P_1$ and $P_2$ and the total dipole moment $P$ as functions of the time. $t_1/U=1$, $n_0=1$, $L=3$ and $\Delta \theta\equiv\theta_1-\theta_2=\pi/2$. (b) Results of supercurrents of planons $J_D$ versus time for various $\Delta\theta$.}
\label{Fig.8}
\end{figure}

\begin{figure}[htbp]
\centering  
\subfigure[]{
\label{jd3}
\includegraphics[width=0.48\columnwidth]{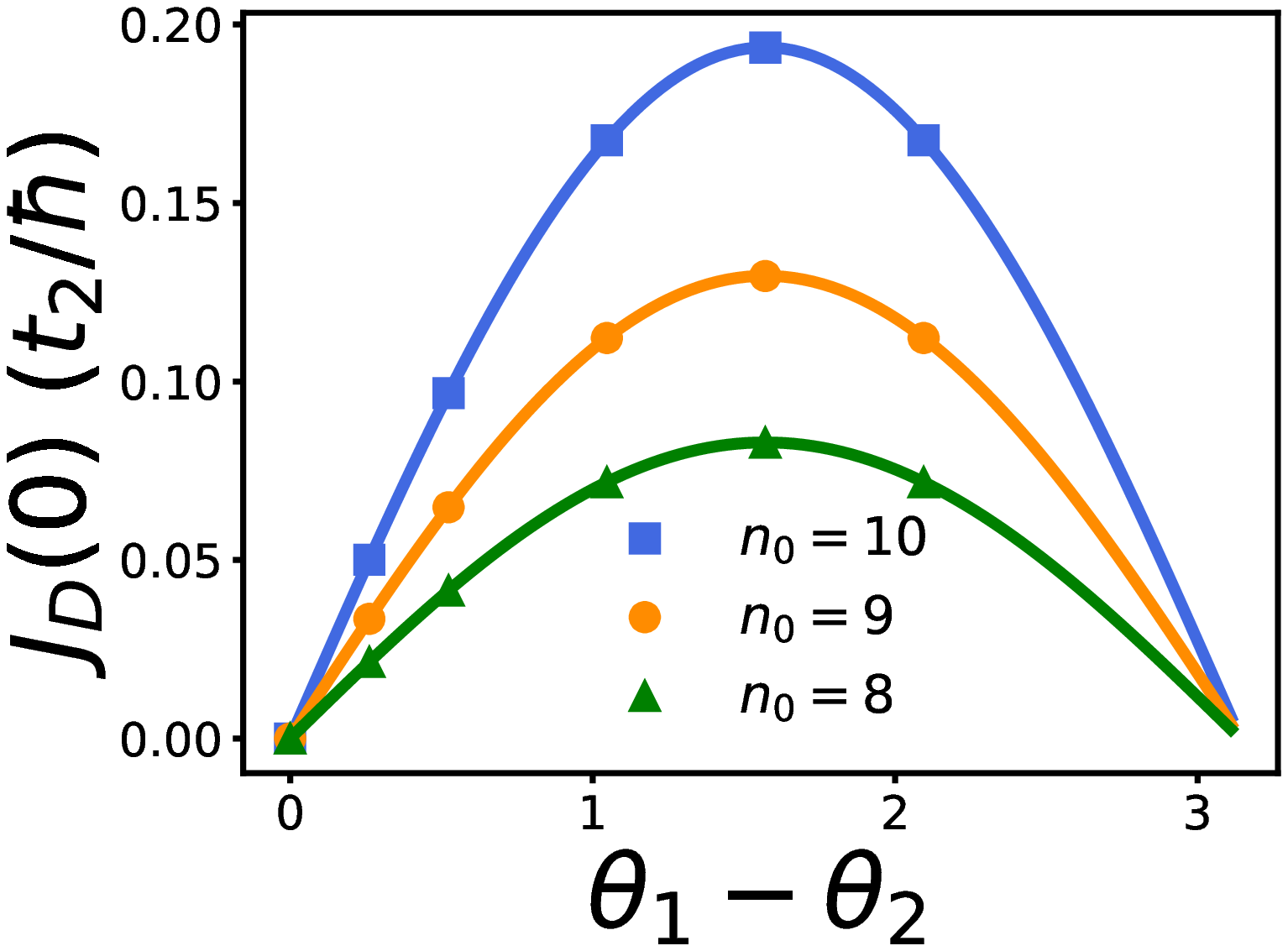}}
\hspace{-5pt}
\subfigure[]{
\label{L}
\includegraphics[width=0.48\columnwidth]{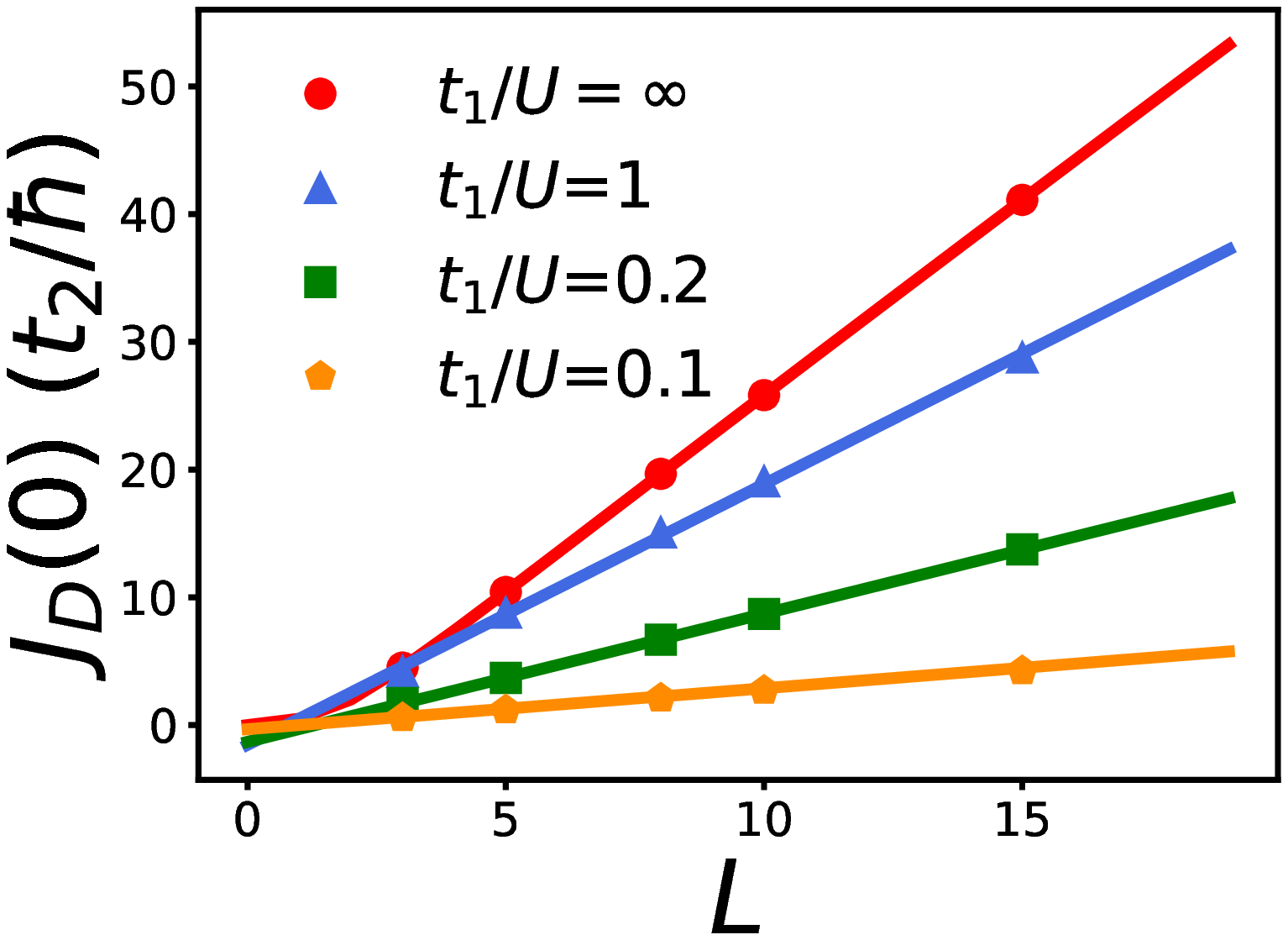}}
\caption{(a) Results of the 
supercurrent of planons at $\tau=0$. $t_1/U$=0.001 and $L=3$. Numerical simulations are shown by dots whereas curves are analytical results from Eq.~(\ref{JJ}) and Eq.~(\ref{A1}). (b) The dependence of $J_D(\tau=0)$ on $L$. $\Delta\theta$ has been chosen as $\pi/2$. When $t_1/U=\infty(0.1)$, numerical results agrees with Eq.~(\ref{A1}) shown by red(orange) lines. The other two lines are linear fits of numerical results. 
}
\label{jd2}
\end{figure}

As shown in Fig.~\ref{dm2}, $P(\tau)$ is a time-independent constant. This is expected, as the kinetic energy in $H_f$ conserves the total dipole moment. However, both $P_1(\tau)$ and $P_2(\tau)$ are time-dependent. We thus define the supercurrent of dipoles as
\begin{equation}
    J_{\text D}(\tau)=\langle\frac{\partial P_1 }{\partial \tau}\rangle=-\langle\frac{\partial P_2 }{\partial \tau}\rangle.
\end{equation}
Fig.~\ref{dc2} shows the result of $J_{\text D}(\tau)$. In later times, $J_{\text D}(\tau)$ becomes a complex function of time, dependent on the competition between $t_2$ and $U$. When $\tau\rightarrow 0$, there exists simple Josephson relation,
\begin{equation}
   J_{\text D}=A\sin{(\theta_1-\theta_2)}\label{JJ},
\end{equation}
where $A$ is a time-independent constant. This can be understood from that the supercurrent of dipoles is written as 
\begin{equation}
\begin{aligned}
    J_{\text D}&=i\langle[H_f,P_1(\tau)]\rangle\\
    &=it_2\sum_j(\langle D^\dagger_{1,j}D_{2,j} \rangle-\langle D^\dagger_{2,j}D_{1,j} \rangle),
\end{aligned}
\end{equation}
where $D_{l,j}=b_{l,j}^\dagger b_{l,j+1}$. When $\tau\rightarrow 0$, the average is taken at the initial state, and $\langle D^\dagger_{1,j}D_{2,j} \rangle=\langle D^\dagger_{1,j}\rangle\langle D_{2,j}\rangle\sim e^{i(\theta_1-\theta_2)}$. $J_D$ is thus given by Eq.~(\ref{JJ}). The constant $A$ can be derived in certain extreme cases,
\begin{equation}
    A=\begin{cases}
        8t_2(\frac{t_1}{U})^2n_0^2(n_0+1)^2(L-1), &\frac{t_1}{U}\ll1,\\
        t_2n_0^2\big[\cos(\frac{2\pi}{L+1})+2\big]\frac{L^2}{L+1}
        , &\frac{t_1}{U}\gg1.
    \end{cases}\label{A1}
\end{equation}
We have made use of the fact that when $t_1/U\ll 1$, the ground state of BHM is given by Eq.(\ref{Short}). In the opposite limit $t_1/U\gg 1$, the ground state can be approximated by a single-particle condensate where all particles in each chain are condensed at the zero momentum state. These analytical results agree with numerical simulations well, as shown in Fig.~\ref{jd2}. Here the tunneling of planons may happen on arbitrary plaquette between the two chains, accounting for the $L$-dependence of $J_D$ as shown in Fig.~\ref{L}.

\begin{figure}[htbp]
\centering  
\includegraphics[width=0.95\columnwidth]{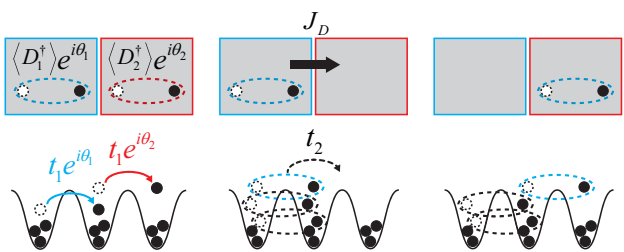}
\caption{ A proposal to observe the dipolar Josephson effect of lineons. In step 1(left), the system is initialized at the ground state of $H_i$ as shown in Eq.~(\ref{3siteHi}). A phase difference $\theta_1-\theta_2$ is imprinted between the dipoles on the left half and right half. In step 2(middle), the Hamiltonian is changed to $H_f$ as shown in Eq.~(\ref{3siteDBHM}). This allows lineons to tunnel but forbids single-particle tunnelings. In step 3(right), the supercurrent of dipoles is measured by tracing the dipole moments of the left and the right half of the system.}
\label{3site}
\end{figure}

Whereas the above discussions apply to planons, Josephson effects also exist for lineons. Fig.~\ref{3site} shows a simple example of a three-site problem. Initially, the Hamiltonian is written as 
\begin{equation}
    H_{i}=-t_1(e^{i\theta_1}b_2^\dagger b_{1}+e^{i\theta_2}b_{3}^\dagger b_2+h.c.)+\frac{U}{2}\sum_{i=1}^3 n_i(n_i-1).\label{3siteHi}
\end{equation}
Although this is a small system with only 3 sites, we could still discuss macroscopic quantum phenomena and dipole condensates in the large $N$ limit, similar to the conventional double-well condensates~\cite{Milburn1997, Leanhardt2004}. Here, the second site divides the whole system into two parts and a phase difference $\Delta\theta\equiv\theta_1-\theta_2$ between the left half and the right half has been imprinted to the system. Since such a finite $\Delta\theta$ exists at the ground state of $H_i$, a supercurrent is absent initially. Now at time $\tau=0$, we change the Hamiltonian to a three-site version of the DBHM
\begin{equation}
    H_f=-t_2(b_2^{\dagger 2} b_3b_{1}+h.c.)+\frac{U}{2}\sum_{i=1}^3 n_i(n_i-1).\label{3siteDBHM}
\end{equation}
Without loss of generality, we define the dipole moment of each half of the system as 
\begin{equation}
\begin{aligned}
        P_L(\tau)&=mn_1(\tau)+\frac{m+1}{2}n_2(\tau),\\
    P_R(\tau)&= \frac{m+1}{2}n_2(\tau)+(m+2)n_3(\tau),
\end{aligned}
\end{equation}where $m$ denotes the coordinate of the first site in an arbitrary reference frame. All following results are independent of $m$ and the choice of the reference frame. The total dipole moment is written as
\begin{equation}
    P(\tau)=\sum_{i=1}^3 (m+i-1) n_i(\tau).
\end{equation}

\begin{figure}[htbp]
\centering  
\subfigure[]{
\label{dm}
\includegraphics[width=0.48\columnwidth]{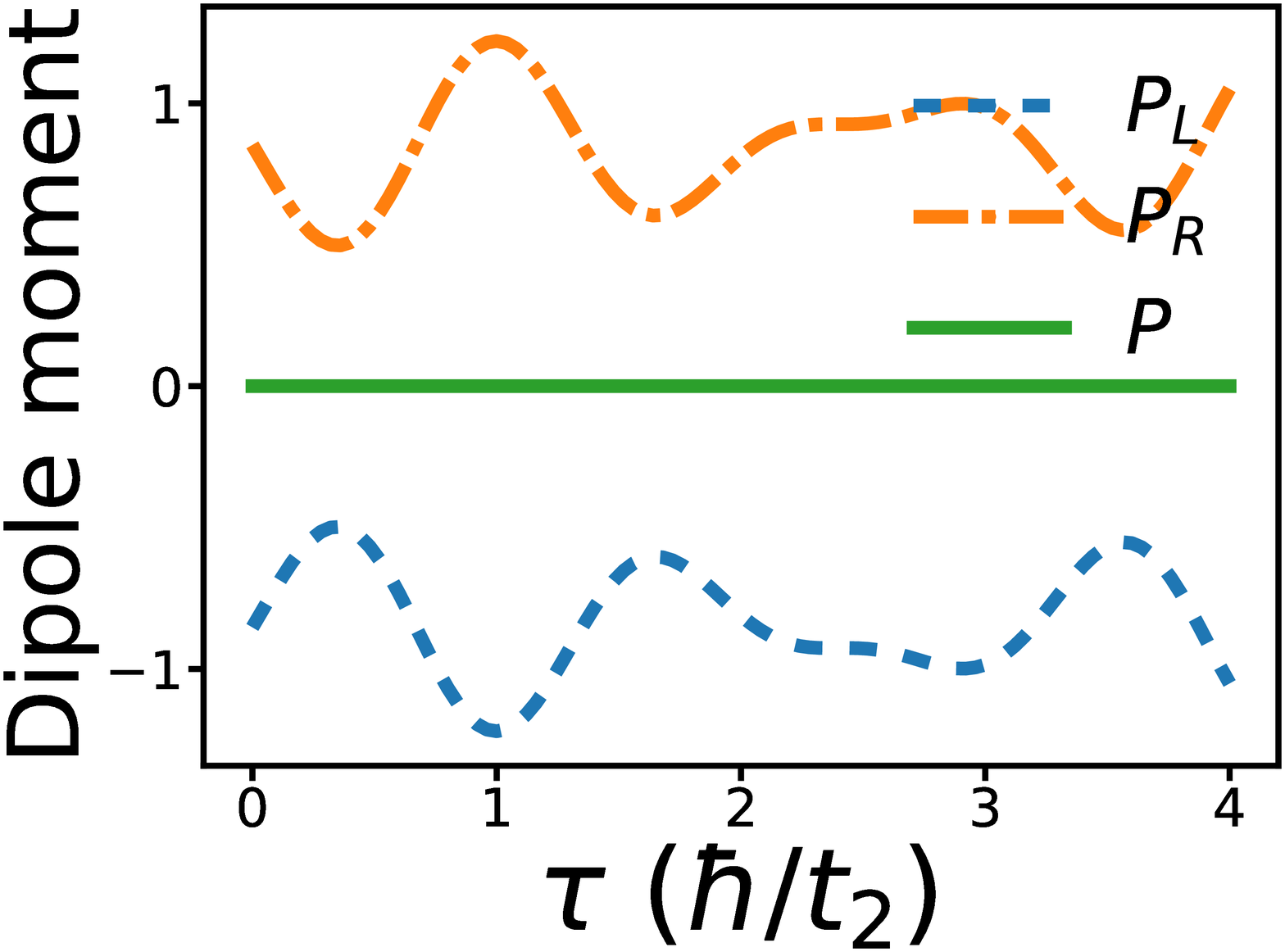}}
\hspace{-5pt}
\subfigure[]{
\label{dc}
\includegraphics[width=0.48\columnwidth]{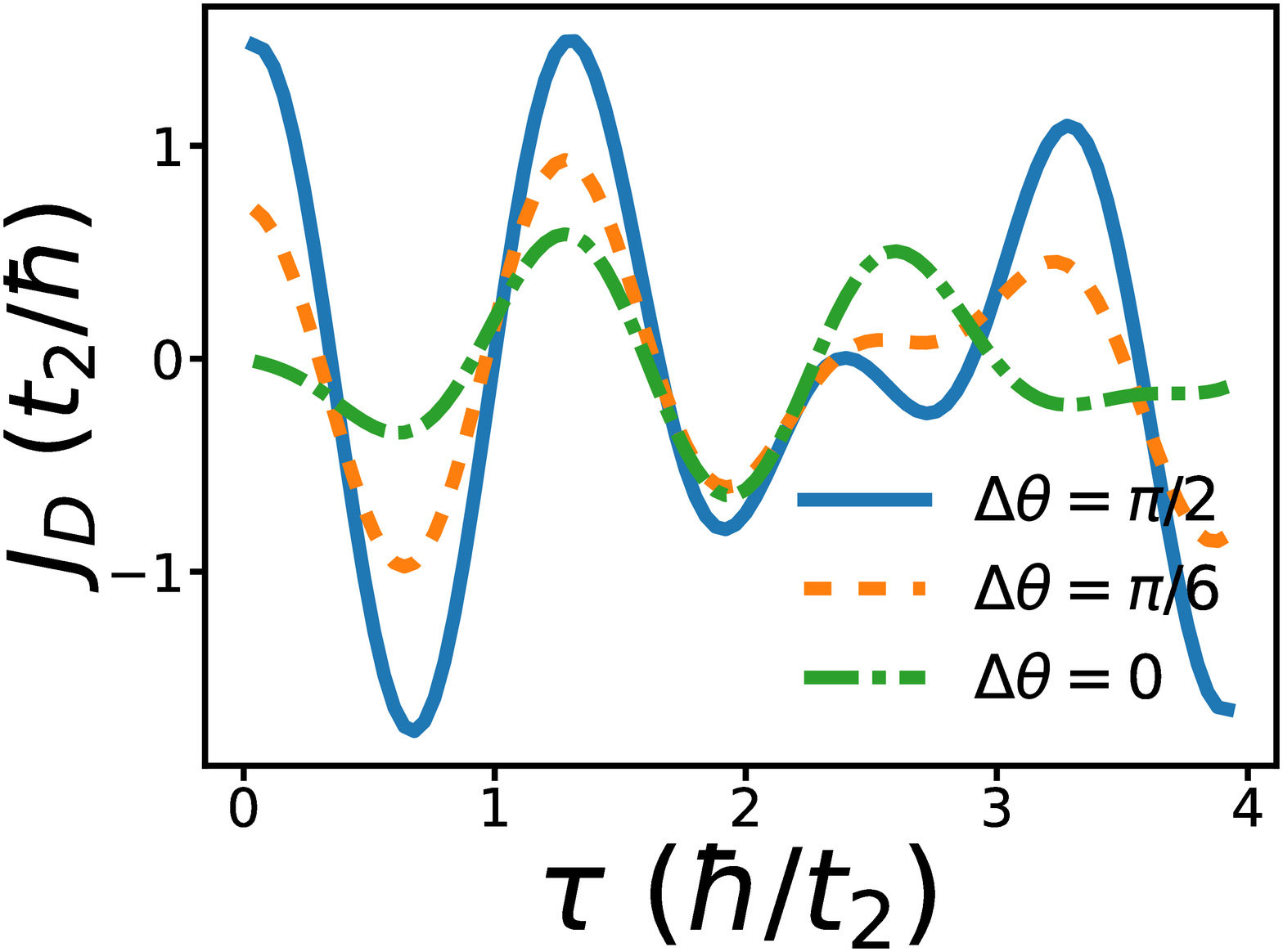}}
\caption{(a) Results of the dipole moment on the left half ($P_L$) and right half ($P_R$) of the system, and the total dipole moment $P$ as functions of the time. $t_1/U=1$, $n_0=1$, $L=3$, and $\Delta \theta=\pi/2$. (b) Results of supercurrents of lineons $J_D$ versus time for various $\Delta\theta$.}
\end{figure}

\begin{figure}[htbp]
\centering  
\includegraphics[width=0.48\columnwidth]{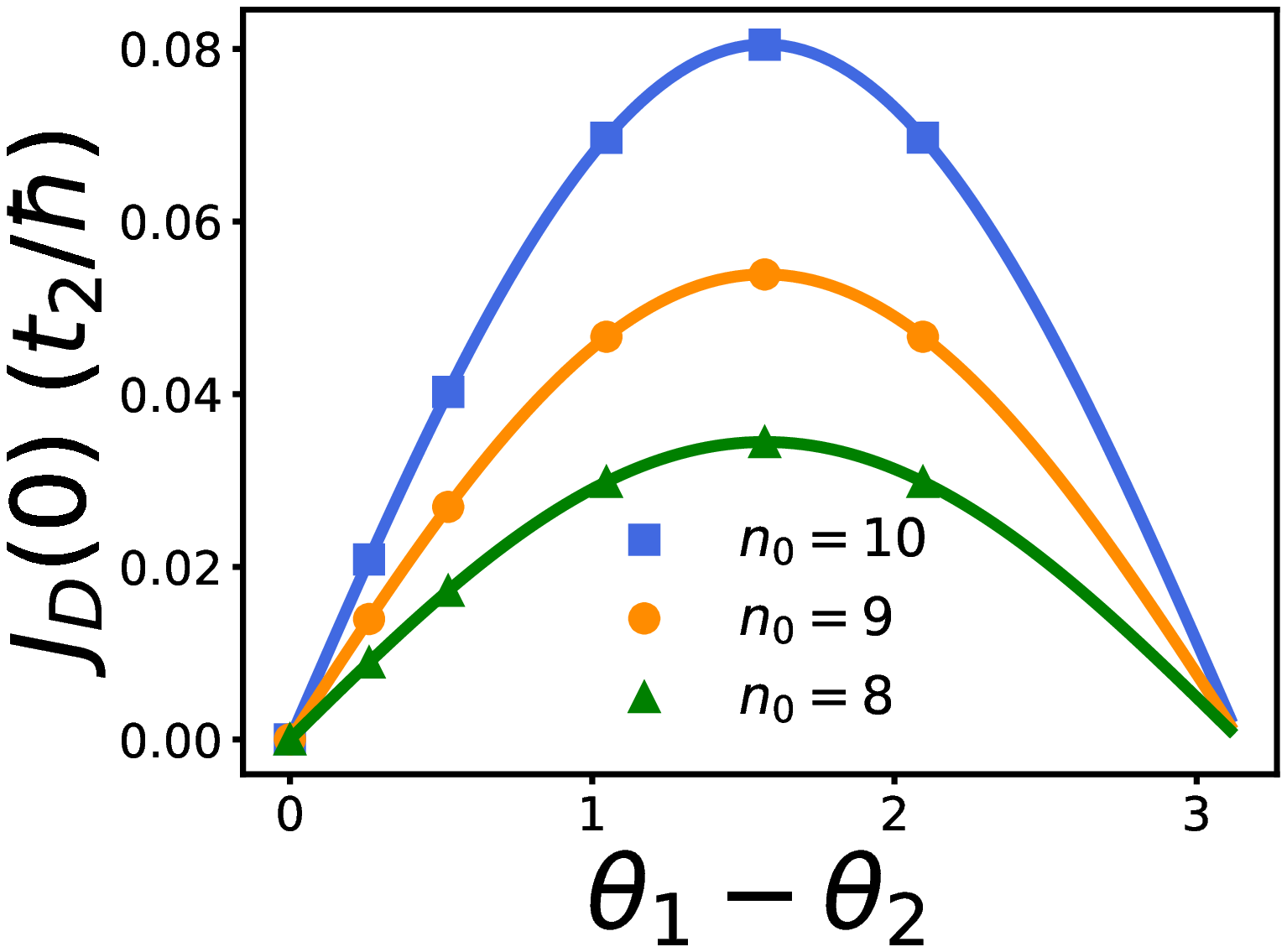}
\hspace{-5pt}
\includegraphics[width=0.48\columnwidth]{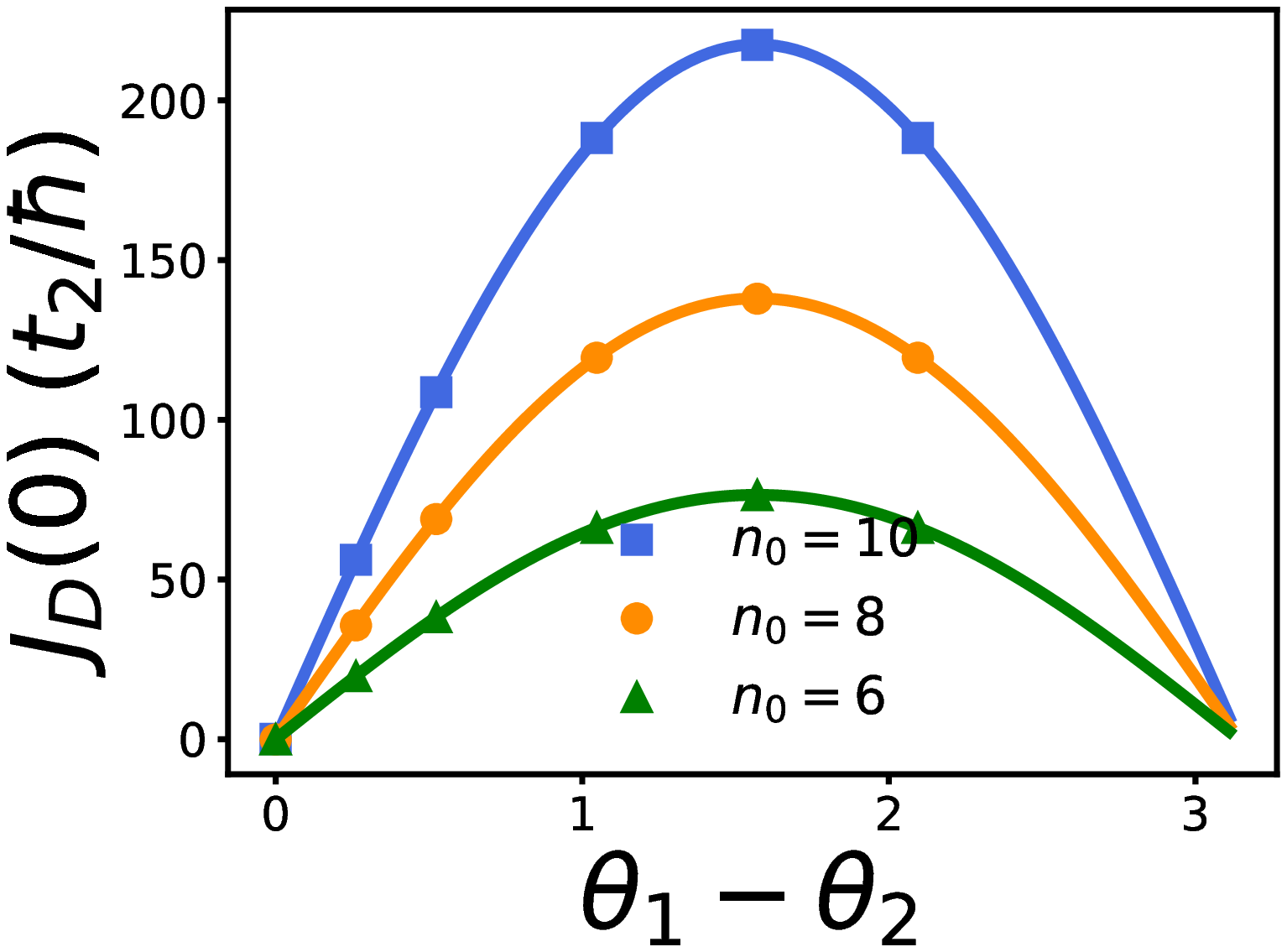}
\caption{Results of $J_D(\tau=0)$ of lineons versus the phase difference at $t_1/U=0.001$(left) and $t_1/U=\infty$(right) in a 3-site system. Numerical results are shown by dots whereas curves are obtained analytically from Eq.~(\ref{JJ}) and Eq.~(\ref{A}). }
\label{jd}
\end{figure}

As shown in Fig.~\ref{dm}, $P(\tau)$ is time-independent as $H_f$ conserves the total dipole moment. From $J_D=\partial P_L/\partial \tau$, we obtain the time-dependent supercurrent of dipoles. Again, Eq.~(\ref{JJ}) applies When $\tau\rightarrow 0$, and the constant $A$ can be derived in certain extreme cases. We have found that
\begin{equation}
    A=\begin{cases}
        \frac{4}{3}t_2(\frac{t_1}{U})^2n_0(n_0+1)(5n_0^2+5n_0-1), &\frac{t_1}{U}\ll1,\\
        \frac{3}{4}t_2(3{n_0}^2-n_0), &\frac{t_1}{U}\gg1.
    \end{cases}\label{A}
\end{equation}
As shown in Fig.~\ref{jd}, the analytical results shown by curves agree well with numerical simulations.


\section{Multipolar condensates}

The above discussions about dipole condensates can be straightforwardly generalized to an arbitrary multipolar condensate. For instance, a quadrupole condensate is readily accessible once the DBHM in Eq.~(\ref{H2}) or Eq.~(\ref{H2p}) is realized. In the limit where $t_2/U=0$, the ground state of a DBHM is $|\text{MI}\rangle$. It has been shown there exists a critical value $(t_2/U)_c$ across which the ground state becomes a dipole condensate~\cite{Senthil2022, Senthil2023}. Though the dipole condensate vanishes when $t_2/U<(t_2/U)_c$, we need to emphasize that the ground state is not featureless. Similar to previous discussions about dipole condensates in BHM, here, a quadruple condensate exists when $t_2/U<(t_2/U)_c$. In the limit $t_2/U\rightarrow 0$, as analogous to Eq.~(\ref{Short}), the ground state of the Hamiltonian in Eq.~(\ref{H2}) is written as
\begin{equation}
    |\Psi_2\rangle\approx|\text{MI}\rangle+\frac{t_2}{3U}\sum_{i} ({b_i^\dagger}^2 b_{i-1} b_{i+1}+h.c.) |{\text{MI}}\rangle.
\end{equation}
As a result of the self-proximity effect, the kinetic energy of the lineon automatically induces a quadrupole condensate on a 1D lattice. 

In 2D, a quadrupole condensate also arises as the ground state of the Hamiltonian in Eq.~(\ref{H2p}) when $t'_2/U\rightarrow 0$,
\begin{equation}
    |\Psi'_2\rangle\approx|\text{MI}\rangle+\frac{t'_2}{2U}\sum_{\bf i} (b_{\bf i+\hat{x}}^\dagger b_{\bf i+\hat{y}}^\dagger b_{\bf i} b_{\bf i+\hat{x}+\hat{y}}+h.c.) |{\text{MI}}\rangle.
\end{equation}
Here the quadrupole condensate is induced by the kinetic energy of the planon, i.e., the ring-exchange interaction. Further controlling the phase of $t'_2$ will allow experimentalists to manipulate the phase of quadrupole condensates and thus deliver the quadrupole Josephson effect. Whereas the movement of a quadrupole could be more complex than a dipole, the generic principle is the same. After producing position-dependent phases, turning on the kinetic energy of quadrupole leads to a supercurrent of quadrupoles.

\begin{figure}[htbp]
    \centering
    \includegraphics[width=0.95\columnwidth]{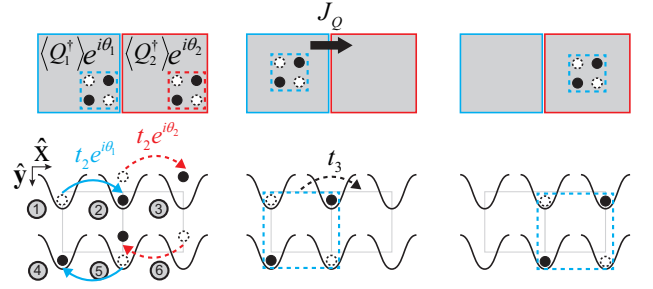}
    \caption{ A proposal to study the quadrupole Josephson effect in experiments. In  step 1(left), the system is initialized at the ground state of the Hamiltonian in Eq.~(\ref{Hqi}). A phase difference $\theta_1-\theta_2$ is imprinted. 
    In step 2(middle),  the Hamiltonian changes to Eq.~(\ref{quadhf}) at $\tau=0$ to allow quadrupoles to tunnel. In step 3(right), the supercurrent of quadrupoles is measured by tracing the quadrupole moment of each half of the system as time goes by.}
    \label{6site}
\end{figure}

Fig.~\ref{6site} shows an example of the quadrupole Josephson effect in a 6-site system. The initial Hamiltonian is written as 
\begin{equation}
\begin{aligned}
    H_i=&-(t_2e^{i\theta_1} b^\dagger_2 b^\dagger_4 b_1 b_5+t_2e^{i\theta_2} b^\dagger_3 b^\dagger_5 b_2 b_6+h.c.)\\
    &+\frac{U}{2}\sum_{\bf i} n_{\bf i}(n_{\bf i}-1).\label{Hqi}
\end{aligned}
\end{equation}
When $t_2/U<(t_2/U)_c$, the ground state of $H_i$ is a quadrupole condensate, and $\theta_1$($\theta_2$) is the phase of quadrupoles on the left(right) plaquette.
At time $\tau=0$, the Hamiltonian is switched to
\begin{equation}
    H_f=-(t_3{b_2^\dagger}^2 b_4^\dagger b_6^\dagger b_5^2 b_1 b_3+h.c.)+\frac{U}{2}\sum_{\bf i} n_{\bf i}(n_{\bf i}-1)\label{quadhf}
\end{equation}that turns on the quadrupole kinetics. As analogous to Eq.~(\ref{JJ}), the supercurrent of  quadrupoles when $\tau\rightarrow 0$ follows the Josephson relation,
\begin{equation}
   J_{\text Q}=A\sin{(\theta_1-\theta_2)}.
\end{equation}

The above discussions apply to any multipolar condensates of arbitrary order. For instance, we consider a quadrupole Bose-Hubbard model, 
\begin{equation}
\begin{aligned}
    H_3=&-\sum_{\bf i}(t_3b^\dagger_{\bf i+ x}b^\dagger_{\bf i+ y}b^\dagger_{\bf i+ z}b^\dagger_{\bf i+ x+y+z}b_{\bf i}b_{\bf i+x+y}b_{\bf i+x+z}b_{\bf i+y+z}\\
    &+h.c.)+\frac{U}{2}\sum_{\bf i} n_{\bf i}(n_{\bf i}-1).
\end{aligned}
\end{equation}
When $t_3/U$ is larger than a critical value, a quadrupole condensate is expected as the ground state. For a generic value of $t_3/U$ less than this critical value, the quadrupole condensate is replaced by octapole condensate. More generically, a multipolar Bose-Hubbard model of the $n$-th order can be written as
\begin{equation}
    H_n=-\sum_{\langle{\bf i},{\bf j}\rangle}t_n M^\dagger_{n\bf{i}} M_{n\bf{j}}+h.c.,
\end{equation}
where $M^\dagger_{n \bf{i}}=\prod_{j=1}^{2^{n-1}}b^\dag_{{\bf i+k}_j}\prod_{j=1}^{2^{n-1}}b_{{\bf i+k}'_j}$ 
is a creation operator of a multipole of the $n$-th order, with a vanishing $m$-th order ($m<n$) multipole moment. 
When $t_n/U$ is smaller than a critical value, a multipolar condensate of $(n+1)$-th order arises as the ground state. Such multipolar Bose-Hubbard models thus provide us with a hierarchy of multipolar condensates. 

\section{Experimental realizations}

Many of our theoretical results are readily accessible using currently available experimental techniques. For instance, the BHM has been widely explored using ultracold atoms in optical lattices~\cite{Zoller1998, Greiner2002, Trotzky2010}. A Mott insulator with short-range correlations has also been observed in experiments~\cite{Bloch2005}. Experimentalists could also increase the temperature to suppress the condensate and achieve a normal phase of bosons at finite temperatures in optical lattices. Therefore, a dipole condensate already exists in laboratories. To further manipulate the phase of a dipole condensate, a phase needs to be added to the tunneling $t_1$. This has also been done in experiments using synthetic gauge fields~\cite{Bloch2013, Ketterle2013}. By tilting an optical lattice, the bare tunneling is suppressed. Applying external lasers to overcome the energy mismatch between the nearest neighbor sites, the laser-induced tunneling naturally inherits the phase of laser, which can also be made position dependent. 

To access the dipolar Josephson effect, correlated pair tunnelings need to be used. The kinetic energy in Eq.~(\ref{H2}) can be been realized using similar schemes of titled optical lattices~\cite{Yang2020, Bloch2021}. In addition, the ring exchange interaction in Eq.~(\ref{H2p}) has also been realized for fermions~\cite{Dai2017} and this scheme can be generalized to bosons. Applying linear field gradients along both directions of a square lattice, single-particle tunnelings are suppressed along both the $x$ and the $y$ directions. However, the ring-exchange interaction satisfies the resonant condition and a pair of bosons can then tunnel simultaneously, producing the ring-exchange interaction as shown in Fig.~\ref{planon}.  

\begin{figure}[htbp]
    \centering
    \includegraphics[width=0.98\columnwidth]{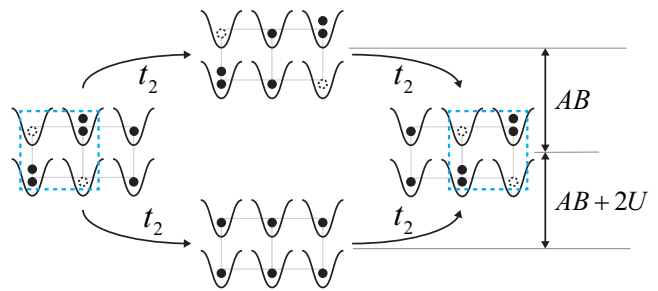}
    \caption{The tunneling of a quadrupole through second-order processes of the ring exchange interaction.}
    \label{t_3fig}
\end{figure}

To design the kinetics of quadruples and higher-order multipoles, extra work is required. 
For instance, we consider a 2D system in Fig.~\ref{6site} and a Hamiltonian in Eq.~\ref{quadhf} with an additional quadratic potential, $(Ax+By)^2/2$. This quadratic potential resembles a quadratic and a linear potential for single particles and dipoles, respectively, and can suppress their tunnelings. A quadruple, however, feels a constant potential and could tunnel without an energy penalty. As shown in Fig.~\ref{t_3fig}, the initial state has a quadruple in the left half of the system and has 
an energy of $AB+2U$ with respect to a Mott insulator with a unit filling. After turning on the ring exchange interaction $t_2$,
two 
pathways of second-order processes allow the quadrupole to move to the right half of the system. The final state has the same energy as the initial one.  
The tunneling amplitude of the quadrupole, $t_3$ in Eq.(\ref{quadhf}), is written as
\begin{equation}
    t_3=\frac{2Ut_2^2}{AB(AB+2U)}.
\end{equation} Whereas the above process represents the nearest neighbor tunneling of the quadrupole, it is straightforward to verify that any long-range tunneling of the quadrupole is suppressed due to destructive interferences between different pathways. Such a scheme could also be generalized to higher dimensions to allow the quadrupole to move in more directions.\\

\section{Summary and outlook} 

In this paper, we have shown that 
dipole condensates are as prevalent as conventional single-particle condensates. 
An ordinary normal phase of bosons, which is defined by a vanishing single-particle order parameter, could readily be a dipole condensate. 
The single-particle kinetic energy automatically  induces dipole condensation. Similarly, the kinetic energy of the $n$-th order multipoles produces a condensation of the $(n+1)$-th order multipoles. This can be understood as 
the self-proximity effect, resembling 
the conventional proximity effect, though no extra external condensates are required. Manipulating the phase of multipolar condensates will allow experimentalists to access a new type of Josephson effect, where the supercurrent of the $n$-th order multipoles arises in the absence of supercurrents of any $m$-th ($m<n$) order multipoles.  

Whereas the dipole condensates and the dipolar Josephson effect are readily accessible in current experiments, it will be useful to realize even higher-order multipolar condensates and higher-order Josephson effects. Though the tunnelings of higher-order multipoles are, in principle, accessible via  higher-order single-particle processes, this requires experimental advancement to realize such tunnelings in laboratories. In addition, it will be interesting to introduce higher rank tensor gauge fields and study their interactions with multipolar condensates.  We hope that our work will stimulate more interest to study multipolar condensates in broader systems including but not limited to fermionic systems and systems with internal degrees of freedom.  

{\bf Acknowledgement} We thank helpful discussions with Senthil Todadri and Ethan Lake. This work is supported by The U.S. Department of Energy, Office of Science through the Quantum Science Center (QSC), a National Quantum Information Science Research Center.

\bibliographystyle{apstest}
\bibliography{dc.bib}

\end{document}